\documentclass{aa}  

\usepackage{graphicx}
\usepackage{newtxtext}
\usepackage{newtxmath}
\usepackage{multirow}
\usepackage{xcolor}

\usepackage{tikz}
\usetikzlibrary{shapes.geometric, arrows}

\usepackage[hidelinks, colorlinks=true, linkcolor=blue, citecolor=blue]{hyperref}

\begin{document}

   \title{The PAU Survey:}

   \subtitle{Enhancing photometric redshift estimation using DEEPz}

   \author
          {I. V. Daza-Perilla \inst{1,2,3} \and 
          M.~Eriksen \inst{4,5} 
          \and
          D.~Navarro-Giron\'es \inst{6,7} \and
          E. J. Gonzalez \inst{3,4}
          \and
          F. Rodriguez \inst{1,3} 
          \and 
          E.~Gazta\~naga \inst{6,7,8}
          \and
          C. M. Baugh \inst{9} 
          \and
          M. Lares \inst{1}
          \and
          L. Cabayol-Garcia \inst{4,5} 
          \and
          F. J. Castander \inst{6,7}
          \and
          M. Siudek \inst{6, 10} 
          \and
          A. Wittje \inst{11} 
          \and
          H. Hildebrandt \inst{11} 
          \and
          R. Casas \inst{6,7}
          \and
          P. Tallada-Crespí \inst{5,12}
          \and
          J. Garcia-Bellido \inst{13}
          \and
          E. Sanchez \inst{12} 
          \and
          I. Sevilla-Noarbe \inst{12}
          \and
          R. Miquel \inst{4,14}
          \and
          C. Padilla \inst{4}
          \and
          P. Renard\inst{15}
          \and
          J. Carretero\inst{5,12}
          \and
          J. De Vicente\inst{12} 
    }

   \institute{
Instituto de Astronom\'ia Te\'orica y Experimental (IATE), CONICET-UNC, C\'ordoba, X5000BGR, Argentina
    \and 
    Facultad de Matem\'atica, Astronom\'ia, F\'isica y Computaci\'on, Universidad Nacional de C\'ordoba (UNC), C\'ordoba, CP:X5000HUA, Argentina
    \and
    Observatorio Astron\'omico de C\'ordoba, Universidad Nacional de C\'ordoba, Laprida 854, C\'ordoba, X5000BGR, Argentina
    \and
    Institut de F\'{i}sica d’Altes Energies (IFAE), The Barcelona Institute of Science and Technology, Campus UAB, 08193 Bellaterra (Barcelona), Spain
    \and
    Port d'Informaci\'{o} Cient\'{i}fica (PIC), Campus UAB, C. Albareda s/n, 08193 Bellaterra (Barcelona), Spain
    \and
    Institute of Space Sciences (ICE, CSIC), Campus UAB, Carrer de Can Magrans, s/n, 08193 Barcelona, Spain
    \and
    Institut d'Estudis Espacials de Catalunya (IEEC), E-08034 Barcelona, Spain
    \and
    Institute of Cosmology \& Gravitation, University of Portsmouth, Dennis Sciama Building, Burnaby Road, Portsmouth PO1 3FX, UK
    \and
    Institute for Computational Cosmology, Department of Physics, South Road, Durham, DH1 3LE, UK
    \and
    Instituto Astrofisica de Canarias, Av. Via Lactea s/n, E38205 La Laguna, Spain
    \and
    Ruhr University Bochum, Faculty of Physics and Astronomy, Astronomical Institute (AIRUB), German Centre for Cosmological Lensing, 44780 Bochum, Germany
    \and
    Centro de Investigaciones Energ\'eticas, Medioambientales y Tecnol\'ogicas (CIEMAT), Avenida Complutense 40, E-28040 (Madrid), Spain
    \and
    Instituto de Fisica Teorica (UAM/CSIC), Nicolas Cabrera 13, Cantoblanco, E-28049 (Madrid), Spain
    \and
    Instituci\'o Catalana de Recerca i Estudis Avan\c{c}ats (ICREA), 0810 Barcelona, Spain
    \and
    Department of Astronomy, Tsinghua University, Beijing 100084, China
}

   \date{Received September 15, 1996; accepted March 16, 1997}

  \abstract{
  We present photometric redshifts for 1\,341\,559 galaxies from the Physics of the Accelerating Universe Survey (PAUS) over 50.38 ${\rm deg}^{2}$ of sky to $i_{\rm AB}=23$. Redshift estimation is performed using {\sc DEEPz}, a deep-learning photometric redshift code. We analyse the photometric redshift precision when varying the photometric and spectroscopic samples. Furthermore, we examine observational and instrumental effects on the precision of the photometric redshifts, and we compare photometric redshift measurements with those obtained using a template method-fitting {\sc BCNz2}. Finally, we examine the use of photometric redshifts in the identification of close galaxy pairs. We find that the combination of samples from W1+W3 in the training of DEEPz significantly enhances the precision of photometric redshifts. This also occurs when we recover narrow band fluxes using broad bands measurements. 
  We show that {\sc DEEPz} determines the redshifts of galaxies in the prevailing spectroscopic catalogue used in the training of {\sc DEEPz} with greater precision. For the faintest galaxies ($i_{\rm AB}=21-23$), we find that {\sc DEEPz} improves over {\sc BCNz2} both in terms of the precision (20-50 per cent smaller scatter) and in returning a smaller outlier fraction in two of the wide fields. 
  The catalogues were tested for the identification of close galaxy pairs, showing that {\sc DEEPz} is effective for the identification of close galaxy pairs for samples with $i_{\rm AB} < 22.5$ and redshift $0.2 < z < 0.6$. In addition, identifying close galaxy pairs common between {\sc DEEPz} and {\sc BCNz2} is a promising approach to improving the purity of the catalogues of these systems.}
    \keywords{galaxies: distances and redshifts -- techniques: photometric --
   methods: data analysis}
   \maketitle
%

\section{Introduction}

The estimation of photometric redshifts is a fundamental task for numerous objectives in cosmology. They enable the mapping of the Universe's large-scale structure and the study of the physical properties of galaxies \citep[e.g.,][]{Jarrett_2004, 2004_Tanaka}. These redshifts also allow the exploration of the Universe's evolution throughout its history, contributing to the validation and improvement of cosmological models, including the study of dark matter and dark energy \citep{Weinberg_2013, Survey2023DESSurveys}. Improving the precision of photometric redshifts is crucial for creating comprehensive catalogues, which in turn enhances the quality and validity of scientific studies based on these redshift data. It also aids in decision-making for observational plans and survey strategies. Achieving high precision in redshifts, for a large number of observations, and over a significant volume of data sampled simultaneously remains a challenge both for spectroscopic and imaging surveys \citep{Hildebrandt_2010, Salvato_2019}.

Spectroscopic surveys are a valuable source of high-precision redshifts; however, this approach  is limited by the number of observations that can be carried out within a specific volume, the time needed to acquire high-quality spectra as well as the challenges associated with studying bright galaxies at low redshifts and faint galaxies at high redshifts. Additionally, selection based on galaxy colour also contributes to these limitations. All these constraints make the generation of spectroscopic catalogues in extensive and uniform volumes difficult. However, the Dark Energy Spectroscopic Instrument (DESI) has created the largest 3D map of the universe ever made and measured the rate at which the universe expanded 8-11 billion years ago with a precision of 1 percent, providing a powerful way to study dark energy  \citep{DESICollaboration2024DESIOscillations}. 

On the other hand, imaging surveys with several filter bands can provide denser galaxy samples over larger volumes by going deeper than spectroscopic surveys, but with lower precision redshifts. The precision of redshifts obtained from a handful of broad bands (BB) is typically $\sim 0.05$, measured from the distribution of relative differences between the spectroscopic redshift, $z_{\rm s}$, and the photometric redshift, $z_{\rm p}$, expressed as $\Delta z \equiv (z_{\rm p} - z_{\rm s})/(1 + z_{\rm s})$ \citep{Hildebrandt_2012, Hoyle_2018}. To enhance the precision in estimating $z_{\rm p}$, it is necessary to have a higher resolution over the wavelength range of the spectral energy distribution (SED). Several surveys have therefore incorporated medium and narrow-band filters to obtain improved photometric redshifts \citep{Marti_2014, Molino_2020}, such as the Advanced Large Homogeneous Area Medium Band Redshift Survey \citep[ALHAMBRA,][]{Molino_2014}, the Javalambre Physics of the Accelerating Universe Survey \citep[J-PAS,][]{Benitez_2014}, the High-redshift and Dead Sources Redshift Survey \citep[SHARDS,][]{Barro_2019}, the Javalambre Photometric Local Universe Survey \citep[J-PLUS,][]{Cenarro_2019}, and the Southern Photometric Local Universe Survey \citep[S-PLUS,][]{De_Oliveira_2019}, as well as the Physics of the Accelerating Universe Survey used here \citep[PAUS,][]{Padilla_2019}.

The PAUS data can be used to obtain high-quality photometric redshift measurements through the  combination of data from 40 narrow-band photometric filters with existing, deeper BB photometry. This combination improves precision compared to estimates based solely on BB \citep{Alarcon_2021}, which has a direct impact on the studies carried out with PAUS, such as measuring intrinsic alignments, galaxy clustering \citep{Johnston_2021}, characterizing properties of galaxies \citep{Tortorelli_2021, Renard_2022, Csizi_2024}, studies over galaxy evolution and formation \citep{Manzoni2024TheRelation}, cosmic shear \citep{Van_den_Busch_2022}, and identifying close galaxy pairs \citep{Gonzales_2023}.

Different techniques have been implemented to improve and expand photometric redshift catalogues estimated using the PAUS data \citep{Eriksen_2019, Alarcon_2020}. \citet{Eriksen_2019} introduced the \textsc{BCNz2} code, which was designed specifically to handle the combination of 40 narrow filters and broadband filters from PAUS and Subaru, obtained as part of the COSMOS-20 survey \citep{Taniguchi_2015}. The \textsc{BCNz2} code fits templates to the observed fluxes and provides $z_{\rm p}$ in the COSMOS field with a high level of precision, which is inferred by comparing to available spectroscopic redshifts. The measured precision is $\sim 0.0037$ for the 50  per cent of all galaxies adjudged to have the highest quality photometric redshift (in terms of the quality factor, see \citealt{Eriksen_2019}) 
with magnitudes $i_{\rm AB} < 22.5$ and measured redshifts
$z_{\rm s} < 1.2$. \citet{Alarcon_2020} extended the development of a hierarchical Bayesian model to estimate redshifts. This method has been tested on realistic simulations, showing that the incorporation of galaxy clustering information improves redshift determinations and reduces systematic redshift uncertainties.

\citet{Eriksen_2020} implemented \textsc{DEEPz}, a deep learning code that includes simulations in the initial training phase and is then trained on observational data. As a result, \textsc{DEEPz} reduced the $\sigma_{68}$ dispersion statistic by 50 per cent at $i_{\rm AB} = 22.5$ compared to existing algorithms in the COSMOS field.

The aim of this work is to determine $z_{\rm p}$ in wide fields of PAUS with {\sc DEEPz}, a method that has not been implemented in these fields. This study also investigates the observational and instrumental effects that may vary between photometric and spectroscopic surveys and the wide PAUS fields on the precision of $z_{\rm p}$ estimates up to apparent magnitudes $i_{\rm AB} = 23$, and to compare our $z_{\rm p}$ measurements with those presented in \citet{Navarro_2023} using the \textsc{BCNz2} method, as both methodologies are affected by different issues. In the case of template methods, it is expected that $z_{\rm p}$ catalogues will exhibit significant dispersion and low precision at high redshifts, whereas machine-learning based methods are occasionally affected by oversampling. Additionally, we investigate whether these catalogues can be implemented in studies such as the identification of close galaxy pairs.


The article is organised as follows. Section~\ref{sec:data} provides an overview of the observational and simulated data used in each field. Section~\ref{sec:methodology}
introduces the \textsc{DEEPz} model used for determination of the $z_{\rm p}$ catalogues and outlines the metrics used to evaluate the $z_{\rm p}$ performance. In this section, we also present the description of the various tests conducted to generate accurate photometric redshifts in three observed wide-fields. Sect.~\ref{sec:catalogue} we present the obtained  $z_{\rm p}$ for each field and the combination of them.  We analyse the variation in the precision of $z_{\rm p}$ according to instrumental and observational effects, and the spectroscopic sample used in the training of {\sc DEEPz}. Additionally, it assesses the performance of $z_{\rm p}$ in identifying galaxy pairs and compares the results obtained with \textsc{DEEPz} and \textsc{BCNz2} and we describe the generation of the catalogues in each field. Sect.~\ref{close_galaxy_pairs} we implement the $z_{\rm p}$ in the identification of close galaxy pairs. Finally, Section~\ref{sec:conclusion} summarises the conclusions of this work.

Throughout, we adopt a Planck 2015 cosmology \citep{Ade_2015} with the following parameters: $H_{\rm 0} = 67.3 \ \text{km s}^{-1} \ \text{Mpc}^{-1}$, $\Omega_{\rm m}=0.315$, and $\Omega_{\rm \Lambda}=0.685$.

\section{Data} \label{sec:data}

Here, we calculate photometric redshift values in three wide fields observed by PAUS, labelled as W1, W3, and G09 (see fig.~\ref{fig:aitoff}). To estimate $z_{\rm p}$ in each field, we used a combination of simulated and observational data. The observational data is comprised of individual exposures, coadded fluxes from 40 narrow bands of PAUS (NB), and spectroscopic redshifts from various surveys. The specific data used depends on the area under consideration, as set out below.

\subsection{PAUS data} 
\begin{figure}
    \centering
    \includegraphics[width=0.9\linewidth]{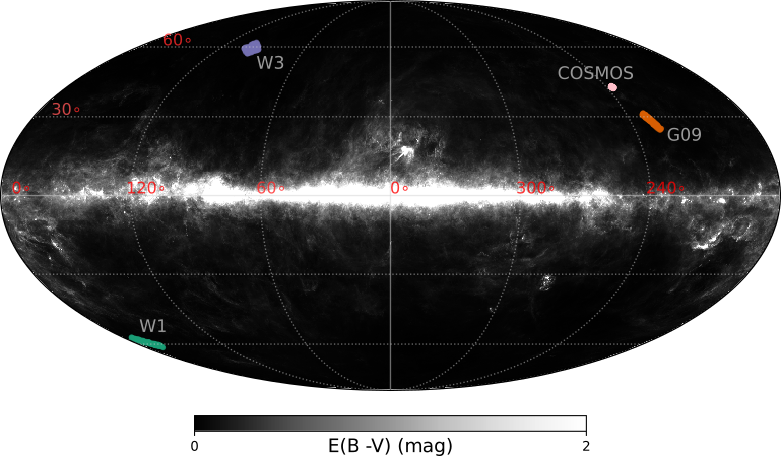}
    \caption{Aitoff Galactic Projection, showing the location of the PAUS fields and the Galactic extinction. The level of extinction is shown in terms of the $E(B-V)$ parameter on a linear scale, as shown by the key. The data are from the dust map provided by the Planck Collaboration \citep{Aghanim_2016}. 
    The W1, G09, W3 and COSMOS wide-fields are shown in green, orange, violet, and pink, respectively.}
    \label{fig:aitoff}
\end{figure}

The PAUS catalogues have been meticulously crafted by the Port d'Informació Cientifica (PIC) data centre. The catalogues are derived from 40 NB optical images, which were acquired with the PAUCam instrument on the William Herschel Telescope at the Observatorio del Roque de los Muchachos in La Palma, Canary Islands \citep{Padilla_2019, Castander_2012}. The data spans the wavelength range from 4500 \r{A} to 8500 \r{A}, with a uniform spacing of 100 \r{A} between contiguous bands, giving an average spectral resolution of $R \sim$ 65. The target fields covered by PAUS are the COSMOS field, the Canada-France-Hawaii Telescope Legacy Survey CFHTLS fields (W1, W3 and W4), and the KiDS/GAMA G09 field  \citep[][]{Heymans2012CFHTLenS:Survey, Erben_2013, Tortorelli_2021}. PAUS has covered 12.04 ${\rm deg}^{2}$ in W1, 15.7 ${\rm deg}^{2}$ in G09, 22.64 ${\rm deg}^{2}$ in W3, and 1 ${\rm deg}^{2}$ in the COSMOS field. PAUS has performed only a few observations in the W4 field, so we have not included this here. Fig.~\ref{fig:aitoff} shows, in the Galactic Aitoff projection, the location of these four fields, along with the dust distribution provided by Planck \citep{Aghanim_2016}. 

The data reduction and galaxy photometry are obtained through two pipelines, the primary pipeline, referred to as the nightly pipeline, which handles all the image processing, with customised algorithms used for photometric calibration \citep{Castander2024TheImages, Serrano_2022}. The multi-epoch and multi-band analysis (\textsc{MEMBA}) pipeline performs forced photometry on a reference catalogue to optimise photometric redshift performance. 


\subsection{W1 Field}

The CFHTLS-WIDE observations cover four patches of the sky at high Galactic latitudes, and are made up of pointings of 1$^{\circ}\times1^{\circ}$ \citep{Erben_2013}. The W1 field has 72 pointings centred on ${\rm RA} =02^{\rm h}18^{\rm m}00^{\rm s}$, ${\rm DEC} 
 =-07^{\rm d}00^{\rm m}00^{\rm s}$. In this field, the PAUS coadded fluxes of 40 NBs correspond to the \textsc{MEMBA} production 1015 and are produced using inverse variance weighting of the individual measurements. Some of these sources have several individual exposures in a single band reaching a maximum of 11 observations. However, most of them have only one observation. We used the CFHTLenS data set for the BB in five filters $u^*$ $g'$ $r'$ $i'$ $z'$ for sources with star\_flag $= 0$. 

The spectroscopic redshift sample used to train \textsc{DEEPz} in this field is made up of data from 13 catalogues. Approximately 60\% of the spectroscopic data used is derived from the VIMOS Public Extragalactic Redshift Survey \citep[VIPERS,][]{Guzzo_2014, Garilli_2014}. VIPERS covers a total area of approximately $24 {\rm deg}^{2}$ located within the CFHTLS-Wide W1 and W4 fields and is limited to objects with magnitudes $i_{\rm AB} < 22.5$ and redshifts in the range of $0.5 < z < 1.5$. The other half of the spectroscopic data, which we will refer to as GAMA-SDSS from now on, comes from various sources, with the Galaxy and Mass Assembly GAMA \citep[][]{Driver_2009, Driver_2011, Baldry_2018} catalogue being one of the most significant contributors. GAMA has obtained redshifts for a quarter of a million galaxies, primarily using the 2dF/AAOmega instrument on the Anglo-Australian Telescope. The final data are described in  \citet{Liske_2015}, with GAMA achieving a high completeness (better than 98\%) down to $r_{\rm AB} < 19.8$ over an area of 180 ${\rm deg}^{2}$ in the equatorial G09, G12, and G15 fields (60 ${\rm deg}^{2}$ each). The W1 field represents 20 per cent of the total sample. The Sloan Digital Sky Survey Data Release 16 SDSS-DR16 \citep[][]{Ahumada_2020} contributes approximately 13 per cent of the dataset, and the following catalogues provide the remaining information: VANDELS \citep[][]{Garilli_2021}, 
KBSS-MOSFIRE \citep[][]{Steidel_2014}, 
VVDS \citep[][]{Le_Fevre_2013}, 
DES\_AAOMEGA \citep[][]{Childress_2017},
3DHST \citep[][]{Brammer_2012},
DES\_IMACS \citep[][]{Dressler_2011},
ZFIRE \citep[][]{Nanayakkara_2016},
C3R2 \citep[][]{Masters_2019}, 
and CDB \citep[][]{Sullivan_2010}. 

\subsection{W3 Field}

The W3 field is centred on ${\rm RA} =14^{\rm h}17^{\rm m}54^{\rm s}$, ${\rm DEC} =+01^{\rm d}19^{\rm m}00^{\rm s}$, with 49 pointings covering a total area of approximately 30 ${\rm deg}^{2}$. Similar to the W1 field, we utilise information from the 40 PAUS NBs (production 1012), along with the five CFHTLenS BB.

In this field, we use a total of seven spectroscopic catalogues to provide the training data. Half of the data comes from SDSS-DR16 \citep{Ahumada_2020}. The bulk of the remaining data (40 per cent) is from DEEP2-DR4 \citep[][]{Newman_2013}, with a magnitude limit of $r_{\rm AB}=24.1$, reaching a $z=1.5$. The remaining minor contributions of spectroscopic information come from 
3DHST \citep{Brammer_2012},
C3R2 \citep{Masters_2019},
CDB \citep{Sullivan_2010}, and
SAGA \citep{Geha_2017}.

\subsection{G09 Field}

The PAUS team conducted observations in a limited portion of the CFHTLS-W2 patch, focusing on 33 pointings centred on the coordinates RA $08^{\rm h}54^{\rm m}00^{\rm s}$ and DEC $-04^{\rm d}15^{\rm m}00^{\rm s}$. This area is located within one of the four equatorial regions mapped by the GAMA survey, namely, 9h (G09), which has been documented in various publications \citep{Driver_2011, Hopkins_2013, Liske_2015, Driver_2016a}. This region is 12 $\times$ 4 ${\rm deg}^{2}$ and is located within the footprint of the public ESO Kilo-Degree Survey and the VISTA Infrared Kilo-Degree Galaxy Survey, commonly known as KiDS and VIKING, respectively \citep[][]{deJong_2013, Kuijken_2015, deJong_2015, Edge_2013, Venemans_2015, Bellstedt_2020}.

KiDS has obtained optical images over 1500 ${\rm deg}^{2}$, captured using OmegaCAM on the VLT Survey Telescope (VST) in the $u$, $g$, $r$, and $i$ filters. Additionally, VIKING has near-infrared coverage through the VISTA telescope in the $Z$, $Y$, $J$, $H$, and $K_{\rm s}$ bands. Consequently, galaxies in the G09 field have information in nine BB.

The GAMA survey is the primary source of spectroscopic information in this area, contributing 80 per cent of the total. The remaining spectroscopic information comes from SDSS. The objects in this field that form the training sample reach apparent magnitudes of $i_{\rm AB}=22$.

A complementary sample called KiDz-COSMOS,  extracted from the KiDS DR5
(Wright et al., in press), has been created, containing information in the nine KiDS and VIKING bands, reaching an apparent magnitude of $i_{\rm AB}=22.5$. Most of the spectroscopic information in this sample comes from the extensive redshift study known as zCOSMOS-bright \citep{lilly_2009}, which covers $1.7 {\rm deg}^{2}$ down to $i_{\rm AB}=22.5$ and $0.1 < z < 1.2$ and is a sub-sample of G10-COSMOS \citep{Davies_2015}.

\subsection{Subsamples}

To evaluate the precision of the $z_{\rm p}$ determined by {\sc DEEPz}, which incorporates supervised machine learning models, we have divided the galaxies with spectroscopic information into three random groups. One of them is used to train the models (the training set), another group is used to evaluate the model performance (the validation set), and a third (the test set) is not used in previous steps, but is reserved to estimate how precise the $z_{\rm p}$ will be for galaxies lacking spectroscopic information. 

In Table~\ref{tab:sets} we show the size of the sets, i.e. the total sample and the training, validation and test sets. The first four columns are the number of galaxies in each field. The last two columns present the data augmentation, specifically the training, validation, and test samples are combined based on their similar photometric characteristics. This augmentation was applied to the W1 and W3 fields from the CFHTLenS survey, which includes (BB) photometry, as well as to a combined dataset from the G09 and COSMOS fields, which integrates BB information from the KiDS and VIKING surveys.

\begin{table*}
    \caption{The first row gives the total sample size; the second row gives the size of the training sets; and the third and fourth rows give the size of the validation and test sets. The columns correspond to the fields. The samples resulting from the union of the training, validation, and test samples of W1 and W3 is presented in column 6, while the union of the samples of G09 and COSMOS is presented in the last column.}
    \centering
    \begin{tabular}{lrrrrr r rr}
    \hline
    Samples                 & \multicolumn{7}{c}{ } \\
                   &  W1       &   W3      & G09    & COSMOS & & W1+W3  & G09+COSMOS\\
    \hline
    \hline
    Total          & 30\,065  & 12\,234    & 4\,313 & 11\,890 & & 42\,299    & 16\,203\\
    Training       & 16\,836   & 6\,850    & 2\,415 & 6\,658  & & 23\,686     & 9\,073\\
    Validation     & 4\,209   & 1\,713    & 604    & 1\,665  & & 5\,922     & 2\,269\\
    Test           & 9\,020    & 3\,671    & 1\,294   & 3\,567  & & 12\,691    & 4\,861\\
    \hline
    \end{tabular}
    \label{tab:sets}
\end{table*}

\subsection{Simulations}

The estimation of photometric redshifts by \textsc{DEEPz} uses a combination of observational and simuated data, as explained in the next section \citep{Eriksen_2020}.
Reproducing the methodology of \textsc{DEEPz}, we have included the code for simulating galaxy spectral energy distributions from flexible stellar population synthesis (FSPS) \citep{Conroy_2009, Conroy_2010} in the neural network training before the input of observational data. This code uses stellar population synthesis (SPS) models to estimate stellar masses, mean ages, metallicities, and star formation histories for different star formation histories. The simulations are generated using the same parameter ranges described in Section 3.2.2 of \citet{Eriksen_2020}. Therefore, the samples in the W1 and W3 fields are identical to those used in that work, consisting of a thousand galaxies uniformly selected with redshift information, as well as fluxes in the PAUS NBs and in the BB of the CFHTLenS survey. Regarding the simulation of the same number of galaxies in the G09 field, the same parameter selection is used. However, in this case the \textsc{FSPS} code is extended to include the nine KiDS and VIKING filters \citep{deJong_2013, Kuijken_2015, deJong_2015, Edge_2013, Venemans_2015}, and the redshift range is extended from $1.5$ to $2.1$.

\section{Methodology}\label{sec:methodology}

The precision of studies conducted using a statistical tool like machine learning depends on several factors: the model, the information available, in particular the amount and quality of the data provided to the model. Among the various factors that influence the performance of the model and, therefore, the determination of photometric redshifts, we have studied in each field the impact of refining the data on the photometric redshifts. 
In Sect.~\ref{deepz}, we describe the {\sc DEEPz} code. Then, in Sect.~\ref{metrics}, we specify the metrics used to study the quality of photometric redshifts. Finally, in Sect.~\ref{processing}, we present the tests of data refinement.

\subsection{{\sc DEEPz}}\label{deepz}

{\sc DEEPz} was designed for photometric redshift estimation in the PAU survey, using NB and BB data  \citep{Eriksen_2020}. Its effectiveness has been tested in the COSMOS field using photometry from the 40 NBs of the PAU survey, the BB of Subaru, and the $u$-band of CFHTLenS.

The methodology of {\sc DEEPz} involves implementing an architecture composed of three neural networks: the first two make up an autoencoder and the third is a mixture density network (MDN). 
Fig.~\ref{fig:deepz} shows the network flow diagram. The autoencoder is used to extract features without prior knowledge of the redshift and to improve performance for faint sources. The latent variables encoded by the autoencoder, along with the original input fluxes divided by the target band, are fed into the MDN, where the target band is defined independently for each BB survey. In the case of CFHTLenS, the target band is the $i$-band, while for KiDS, it is the $r$-band. The third neural network, the MDN gives the probability distribution of the photometric redshift. The mode of this distribution is $z_{\rm p}$ for a specific galaxy. In datail, the network architecture consists of an autoencoder with 10 layers and 250 nodes in both the encoder and decoder. Each layer includes linear transformations followed by ReLU non-linearities, batch normalization, and a 2\% dropout, except for the last three layers. The autoencoder is fed galaxy flux ratios. The $z_{\rm p}$ network, which follows the same structure as the autoencoder, takes both the galaxy flux ratios and autoencoder features as input. It includes 1\% dropout after all linear layers. This network is a MDN, representing the redshift distribution as a linear combination of 10 normal distributions.

\tikzstyle{steps} = [rectangle, minimum width=2cm, minimum height=0.7cm, 
                     text centered, text width=3cm, draw=black, fill=black!20]
\tikzstyle{encoder} = [rectangle, minimum width=2cm, minimum height=0.7cm, 
                     text centered, text width=2cm, draw=black, fill=violet!50]

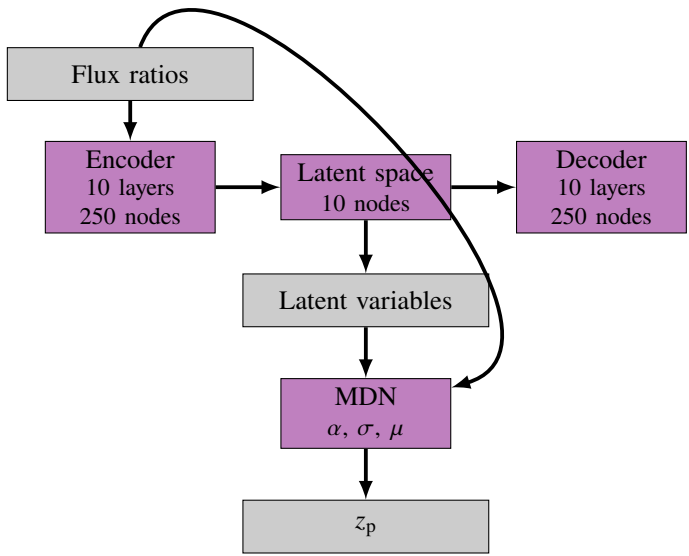
\begin{figure}[ht]
\begin{tikzpicture}[node distance=1.5 cm]
\node (step1) [steps, xshift=1cm, yshift=1cm] {Flux ratios};

\node (encoder) [encoder, below of=step1] {Encoder\\ {\small 10 layers \\ 250 nodes}};
\node (Latent) [encoder, right of=encoder, xshift=1.6cm] {Latent space\\ {\small 10 nodes}};
\node (decoder) [encoder, right of=Latent, xshift=1.6cm] {Decoder\\ {\small 10 layers \\ 250 nodes}};

\node (features) [steps, below of=Latent] {Latent variables};

\node (mnd) [encoder, below of=features] {MDN \\ {\small $\alpha$, $\sigma$, $\mu$}};
\node (zp) [steps, below of=mnd] {$z_{\rm p}$};

\draw[-latex,line width=0.5mm] (step1) -- (encoder);
\draw[-latex,line width=0.5mm] (encoder) -- (Latent);
\draw[-latex,line width=0.5mm] (Latent) -- (decoder);
\draw[-latex,line width=0.5mm] (Latent) -- (features);
\draw[-latex,line width=0.5mm] (features) -- (mnd);
\draw[-latex,line width=0.5mm] (step1) to[out=70,in=17] (mnd);
\draw[-latex,line width=0.5mm] (mnd) -- (zp);

\end{tikzpicture}
\caption{DEEPz network flow diagram.  The data are shown in grey and the networks in violet. The network architecture consists of an autoencoder with 10 layers and 250 nodes in both the encoder and decoder. Each layer includes linear transformations followed by ReLU non-linearities, batch normalization, and a 2\% dropout, except for the last three layers. The autoencoder is fed galaxy flux ratios. The $z_{\rm p}$ network, which follows the same structure as the autoencoder, takes both the galaxy flux ratios and autoencoder features as input. It includes 1\% dropout after all linear layers. This network is a MDN, representing the redshift distribution as a linear combination of 10 normaldistributions.}\label{fig:deepz}
\end{figure}

We use pretraining to help adjust the weights of neural networks so that they can later be fine-tuned with observational data. The model training process begins with simulations. Performing this pre-training step before training with observational data reduces the scatter in $z_{\rm p}$ by 50 per cent for faint sources in terms of apparent magnitude. The subsequent training is conducted with observational data, including data augmentation by constructing several coadd fluxes from individual NB exposures. These are generated on-the-fly and weighted inversely by variance, including each exposure with a probability $\alpha$, which is set here to $\alpha = 0.8$. This causes each galaxy to appear differently to the network in each epoch, i.e., each time the network parameters are adjusted, mimicking repeated measurements over the same sky areas and following systematic patterns to produce combined measurements with reduced noise, allowing the observation of fainter objects. This technique encompasses multiple issues when applied to observational data, such as variability in the number of exposures and the need to inform the network about which measurements are present.

Here, compared with \cite{Eriksen_2020}, we increase the number of layers in the autoencoder from ten to 20 while keeping the number of latent variables unchanged. Additionally, we adapted the input dimensions of the networks to fit the number of bands used. Specifically, in the W1 and W3 fields, the inputs comprise a maximum of 40 NBs and five BB from CFHTlens, unlike the G09 field, where 40 NBs and nine BBs from KiDS and VIKING are used. Additionally, we studied the removal of eight blue NBs and the use of only NBs, as we outline in the Sect.~\ref{processing}.

\subsection{Metrics}\label{metrics}

To assess the precision of $z_{\rm p}$, we have used the $\sigma_{68}$ statistic as the primary metric (see Eq.~\ref{eq:sigma68}). This statistic is calculated from $\Delta z$. Then

\begin{equation}
    \sigma_{68} \equiv \frac{P[84] - P[16]} {2},
    \label{eq:sigma68}
\end{equation}
where $P[x]$ represents the value of the $\Delta z$ distribution at percentile $x$. $\sigma_{68}$ is a centralised measure of the width or scatter in the accuracy of the estimated redshifts that is not affected by outliers. 

We formally define \emph{outliers} as galaxies that meet the condition

\begin{equation}
    \frac{\left|z_{\rm p} - z_{\rm s}\right|} {1 + z_{\rm s}} > 0.02.
    \label{eq:outliers}
\end{equation}
Note that this equation is stricter than the outlier criteria typically used in BB papers \citep{Hildebrandt_2012}. 

\subsection{Data refinement tests}\label{processing}

During the reduction process and calibration, the collected observations undergo a series of steps and techniques aimed at improving their quality and completeness, eliminating possible errors in the measurements or defects such as noise or Galactic extinction, and extracting the most relevant information. These procedures involve curation and preprocessing techniques,
along with the use of statistical tools such as univariate analysis for the selection of relevant features.

Here, seven tests were conducted to examine whether curation and preprocessing procedures, dependent on observational conditions and instruments, affect the precision of $z_{\rm p}$. These tests ranged from combining measurements made by two instruments to selecting specific bands. Each test generated diverse datasets, which influenced the training of {\sc DEEPz} and, consequently, the precision of the photometric redshifts. 

Before describing the tests, we set out the baseline sample first

\begin{itemize}

\item{\it Baseline}: As a starting point for these studies, we exclude galaxies without a measurement in any of the BBs. Consequently, the samples of galaxies, both with and without spectroscopic information, are reduced by 9\%, 11\%, and 2\% in the W1, W3, and G09 fields, respectively. These samples are used as a reference to evaluate and compare the performance of the tests. 
\end{itemize}
\vspace{0.2cm}

\hspace{-0.6cm}
Next we describe each of the tests performed.

\begin{itemize}

\item {\it Cross-validation (CV)}: The W1 and W3 fields have measurements classified as missing in the $i$-band of CFHTLenS due to issues with the filter identified as 'i.MP9702' during observation. However, new measurements were obtained using the successor filter, 'i.MP902', and part of the sample was completed with these values. Therefore, we evaluated the impact on the precision of the redshift by combining the measurements from these two filters, compared to the performance obtained without considering ’i.MP902’ measurements, i.e., with the {\it baseline} sample. This is done to validate that there is no impact on $z_{\rm p}$ when using one of these two filters.

\item {\it Infer missing NB values (non-NaN)}: The catalogue contains missing observations in some NBs. The \textsc{DEEPz} architecture uses a fixed number of inputs, making the code unable to handle a variable number of NBs. In \citet{Eriksen_2020}, the application of the code was restricted to galaxies with measurements in all 40 NBs. Imposing this condition results in the loss of 55.22\% of the galaxies in the W1 field, around 38.52\% in the W3 field, and 57.82\% in the G09 field. 
To reduce the impact on completeness, instead of removing all galaxies, we infer the missing NB values for galaxies with at least 30 NB measurements. In this case, the reduction in the sample size is 30\% of the total sample of galaxies with at least one observation in the W1 field, approximately 28\% in the W3 field, and 42\% in the G09 field. 
The NB estimate was generated through a quadratic fit using the information from the $g$, $r$, and $i$-bands. This type of study allows us to estimate a more complete galaxy catalogue, including galaxies without observations in all filters.

\item {\it Low Signal-to-Noise Ratio Fluxes (Low SNR)}:
We investigate sources in the CFHTLenS catalogue that have at least one BB with a low signal-to-noise ratio, denoted by a value of 99. To generate a catalogue with a greater number of objects, we examine the impact of including sources under these conditions, even though this may not be the case for all bands, and catalogues are usually generated with high S/N based on the target band..

\item {\it Bright Galaxies ($i_{\rm AB} < 22.5$)}:
We study if there is a significant difference in precision when faint galaxies are included or discarded in the analysis. Therefore, we cut the sample of galaxies at an apparent magnitudes of 22.5 according to the target band of the BB surveys corresponding to each wide field.

\item {\it Galactic Extinction Correction (Ext. Correction)}:
Although the wide fields used in this study are not at low Galactic latitudes, they are weakly and differentially affected by the dust, gas, and stellar density of the Milky Way, though the latter does not affect extinction. Therefore, we performed a correction to the NB and BB to undo the effects of Galactic extinction. Before the correction, we removed the Galactic extinction on the BBs of CFHTLenS based on the dust maps presented in \citet{Schlegel_1998}, which already included their apparent magnitudes, unlike the apparent magnitudes of the KiDS and VIKING surveys where they are not extinction corrected in our data. Subsequently, the correction was carried out for the BBs and NBs using the extinction values, $E(B-V)$, provided in the Planck 2015 thermal dust map \citep{Aghanim_2016}. 

The Galactic extinction value at the source position and the corresponding correction factors were estimated through

\begin{equation}
    \phi_{\text{corr}} = \phi_{\text{uncorr}} \cdot \frac{1}{{C_0 \cdot E(B-V)^2 + C_1 \cdot E(B-V) + C_2}} \, , \quad 
    \label{eq:3}
\end{equation}
where the band-dependent coefficients
($C_0$, $C_1$, and $C_2$) were estimated using a second-degree polynomial fit to the median extinction affecting the Pickles stellar templates \citep{Pickles_1998}.
\end{itemize}

We conducted a univariate analysis of the bands and $z_{\rm p}$ using the Pearson correlation, mutual information, and regression methods (see Appendix~\ref{app:appendix_a}). We found that the BBs behave similarly in score to the NBs in terms of their importance for determining $z_{\rm p}$ according to wavelength and these methods, so it is interesting to study the case of only using the NBs to determine $z_{\rm p}$. Additionally, we found a lower score for wavelengths in the blue, so we investigated their importance for the precision of $z_{\rm p}$. The following two tests are based on this study. 

\begin{itemize}

\item {\it Only NBs (NBs)}: We consider eliminating the BBs and using only the NBs. In this case, the NBs are normalised by creating an artificial band from the average of the NBs within the wavelength range of the target selection band, i.e., the $i$-band in the case of fields W1 and W3 and the $r$-band in the case of field G09. This allows us to study whether the more general SED characteristics at low wavelength resolution and high SNR provided by the BBs are important in the precision of $z_{\rm p}$. Therefore, as input, {\sc DEEPz} uses the 40 NB fluxes normalized by the artificial band. This change in the number of bands used to determine $z_{\rm p}$ results in a change in the number of neurons in the input layer of the neural networks.

\item {\it SED in the blue wavelength range (blue-lines)}: 
Since the univariate analysis shows a low correlation for wavelengths in the blue range of the galaxy SED, we explore the possibility of saving time by disregarding the detailed information about the SED in the blue wavelength range, contained in eight filters of the PAU survey, within the wavelength interval of 4550 $\text{\AA}$ to 5250 $\text{\AA}$. Through this test, we evaluate the relevance of lines identified in the blue region with respect to the general SED characteristics within this spectral range when considering all wide bands.
\end{itemize}

In Table~\ref{tab:studies},  we provide a brief description of each test and specify the number of galaxies in the training and validation sets for each case.

\begin{table*}[ht]
\centering
\begin{tabular}{p{4 cm}p{6cm}p{3.2cm}p{3.2cm}}
\hline
\textbf{Data refinement tests} & \textbf{Description} & \textbf{Training Set} & \textbf{Validation Set} \\
\hline
\hline
\multirow{3}{4 cm}{Baseline} & \multirow{3}{6 cm}{Galaxies without measurements in any of the bands are excluded.} & \small{W1+W3:}\hspace{1 cm}22\,378 &  \small{W1+W3:} \hspace{1 cm} 5\,581 \\
&                 & \small{G09+COSMOS: \, 9\,073} & \small{G09+COSMOS: \, 2\,269} \\
& & & \\

\multirow{3}{4 cm}{Cross-validation (CV)} & \multirow{3}{6 cm}{Comparison of $i$-band filters to confirm that the two CFHT filters provide consistent and comparable measurements.} & \small{W1+W3:}\hspace{1 cm}23\,592 &  \small{W1+W3:} \hspace{1 cm} 5\,581 \\
&                 & \small{G09+COSMOS: \, 7\,606} & \small{G09+COSMOS: \, 1\,902} \\
& & & \\
& & & \\
\multirow{3}{4 cm}{Impute (non-NaN)} & \multirow{3}{6 cm}{Investigation into handling galaxies with NaN values in the NB fluxes and the use of interpolation through p{3.2cm}p{3.2cm}a quadratic fit.} & \small{W1+W3:}\hspace{1 cm}30\,383 &  \small{W1+W3:} \hspace{1 cm} 7\,561\\
&                 & \small{G09+COSMOS:}\hspace{0.2 cm}7\,606 & \small{G09+COSMOS:}\hspace{0.2 cm}1\,902  \\
& & & \\
& & & \\

\multirow{3}{4 cm}{Low signal-to-noise ratio fluxes (Low SNR)} & 
\multirow{3}{6 cm}{Investigation of the treatment of low signal-to-noise ratio fluxes including those sources with magnitudes equal to 99.} & \small{W1+W3:}\hspace{1 cm}22\,118 &  \small{W1+W3:} \hspace{1 cm} 5\,529 \\
&                 & \small{G09+COSMOS:}\hspace{0.2 cm}7\,606 & \small{G09+COSMOS:}\hspace{0.2 cm}1\,902 \\
& & & \\
& & & \\
\multirow{3}{4 cm}{Bright galaxies ($i$ $<$ 22.5)} & 
\multirow{3}{4 cm}{We limit the magnitudes of the selected band to values less than 22.5.} & \small{W1+W3:}\hspace{1 cm}22\,378 &  \small{W1+W3:} \hspace{1 cm} 5\,581 \\
&                 & \small{G09+COSMOS:}\hspace{0.2 cm}7\,606 & \small{G09+COSMOS:}\hspace{0.2 cm}1\,902 \\
& & & \\
& & & \\
\multirow{3}{4 cm}{Galactic extinction correction (Ext. Correction)} & 
\multirow{3}{4 cm}{Study on the Galactic extinction correction in the BB and NB fluxes.}  & \small{W1+W3:}\hspace{1 cm}22\,378 &  \small{W1+W3:} \hspace{1 cm} 5\,581 \\
&                 & \small{G09+COSMOS:}\hspace{0.2 cm}7\,606 & \small{G09+COSMOS:}\hspace{0.2 cm}1\,902 \\
& & & \\
& & & \\
\multirow{3}{4 cm}{Only NBs (NBs)} & \multirow{3}{6 cm}{Analysis of the importance of the more general features of the SED provided in the BBs.} & \small{W1+W3:}\hspace{1 cm}22\,378 &  \small{W1+W3:} \hspace{1 cm} 5\,581 \\
&                 & \small{G09+COSMOS:}\hspace{0.2 cm}7\,606 & \small{G09+COSMOS:}\hspace{0.2 cm}1\,902 \\
& & & \\
& & & \\

\multirow{3}{4 cm}{SED in the blue wavelength range (blue-lines)} & 
\multirow{3}{6 cm}{Indicates the importance of the lines detected in the blue range.}  & \small{W1+W3:}\hspace{1 cm}22\,378 &  \small{W1+W3:} \hspace{1 cm} 5\,581 \\
&                 & \small{G09+COSMOS:}\hspace{0.2 cm}7\,606 & \small{G09+COSMOS:}\hspace{0.2 cm}1\,902 \\
& & & \\
\hline
\end{tabular}
\caption{Summary of studies on observational effects in photometric redshift estimation.  Dimensions of the training and validation sets for the combination of data in W1 and W3, and for the combination of data in the fields G09 and COSMOS.}
\label{tab:studies}
\end{table*}


\section{Results and Discussion}\label{sec:catalogue}

In Sect.~\ref{subsec:Photo_z_precision}, we perform a comprehensive analysis of the results obtained in each wide field and their combination. Additionally, we analyse the outcomes corresponding to the data refinement tests described in Sect.~\ref{processing}. In Sect.~\ref{subsec:effect_of_spectroscopic_surveys} we consider the origin of the spectroscopic redshifts and the effect on the $z_{\rm p}$.
Additionally, in Sect.~\ref{subsec:deepzVSbcnz} we compare our best results with the $z_{\rm p}$ obtained through the {\sc BCNz2} method. Finally, we describe the generation of the $z_{\rm p}$ catalogue in the wide fields in Sect.~\ref{subsec:catalogues}. 

All measurements are performed in the intervals $18 \leq i_{\rm AB} \leq 23$ and $z_{\rm s} \leq 2$ unless stated otherwise.

\subsection{Photometric redshift precision}\label{subsec:Photo_z_precision}


In our study, the precision of the $z_{\rm p}$ is intrinsically linked to the field under consideration and the data available in each field. This is because the latter are used in the training process and validation, which depends on the fluxes of the NBs and BBs, the number of examples, and the distributions of $z_{\rm s}$.

Therefore, in this section, we present the comparison of the results using the baseline samples of each field and their combinations. In Appendix~\ref{app:appendix_b}, the results of each data refinement test in each field and their combination are shown, implementing the training and validation samples differentiated by these conditions.

\vspace{0.2 cm}

\subsubsection{Performance by field}

Initially, we consider each field individually, as well as the combination of the CFHTLenS fields, W1 with W3, and the KiDS /VIKING fields G09 with COSMOS. The analysis is carried out in terms of the $\sigma_{68}$ trends, which are analysed in relation to the $i$-band magnitude limit and $z_{\rm s}$. 

\begin{figure}
    \centering
    \includegraphics[width=1\linewidth]{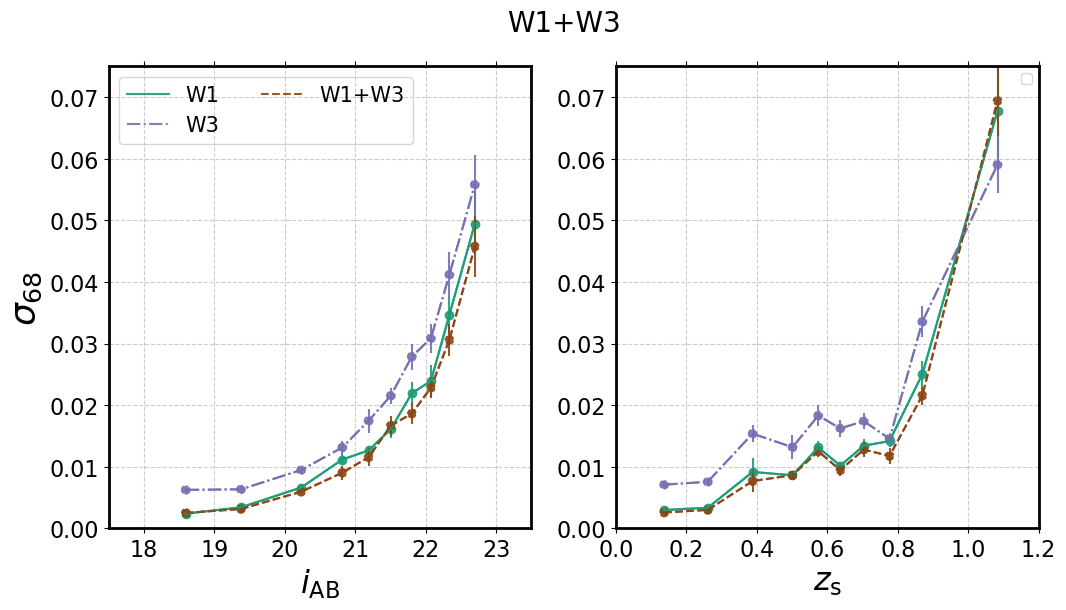}
    \includegraphics[width=1\linewidth]{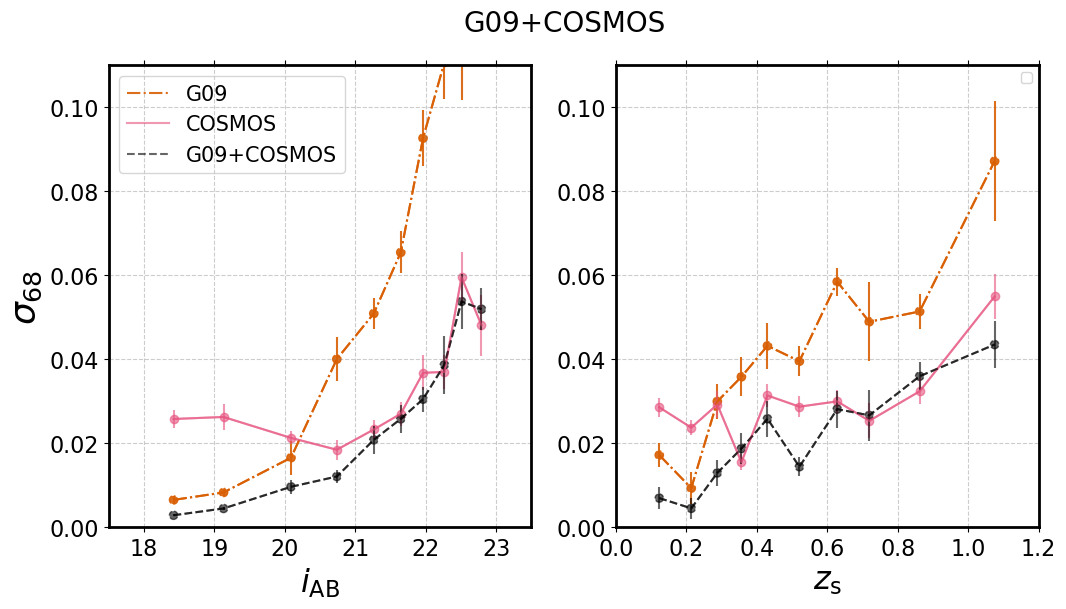}
    \caption{Trends in the measurement of $\sigma_{68}$. The distribution is divided into ten bins, each containing the same number of objects, based on the apparent magnitude in the $i$-band and $z_{\rm s}$, in the left and right columns, respectively. These trends show the precision, in terms of the centralised scatter $\sigma_{68}$, in the W1+W3 and G09+COSMOS validation sets.}
    \label{fig:refinement_data_a}
\end{figure}

We compare the precision of the baseline samples in the W1 and W3 fields and the combination W1+W3. We performed the same analysis for the baseline samples in the G09 and COSMOS fields and their combination. The behaviour of these two cases is shown in Fig.~\ref{fig:refinement_data_a}.

As expected, in all fields, it can be seen that the precision in terms of $\sigma_{68}$, as a function of $i$ and $z_{\rm s}$, is higher for bright and nearby galaxies (lower $\sigma_{68}$), and decreases for faint and distant galaxies (higher $\sigma_{68}$).  When comparing the values obtained from each test in each individual field, we find a systematic superiority in the precision of W1 over W3 and of G09 over COSMOS. We also find that the combination of fields with the same NBs and BBs, W1+W3 and G09+COSMOS, produces better results than the individual samples W3 and COSMOS, respectively, in terms of the $\sigma_{68}$ trends when studied as a function of $i$ and $z_{\rm s}$, compared to using individual samples from each field. Regarding the comparison of results between the W1+W3 and G09+COSMOS samples, differentiated by the  photometry and spectroscopic redshift catalogues, we observe that in all studies, W1+W3 outperforms G09+COSMOS. The G09 and COSMOS fields have four more BBs than the CFHTLenS fields but with lower SNR; this does not generate differences in the trends of the metrics measured as a function of $i$ and $z_{\rm s}$ between G09+COSMOS and the CFHTLenS fields. However, we note that {\sc DEEPz} determines $z_{\rm p}$ with less precision in G09+COSMOS compared to W1+W3 as in the individual comparisons.

\subsubsection{Analysis of the data refinement tests}

\begin{figure*}
    \centering
    \includegraphics[width=0.80\linewidth]{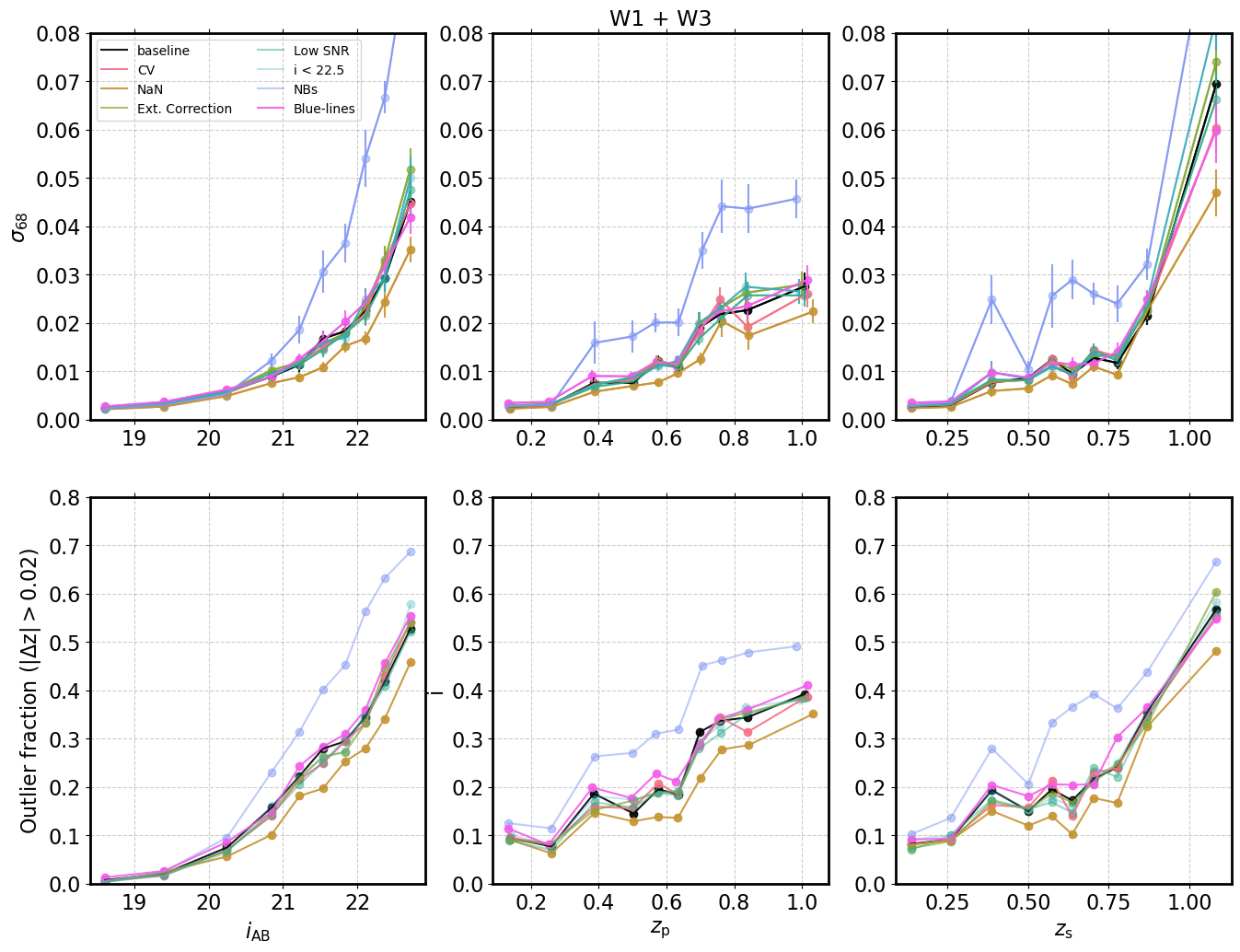}
    \includegraphics[width=0.80\linewidth]{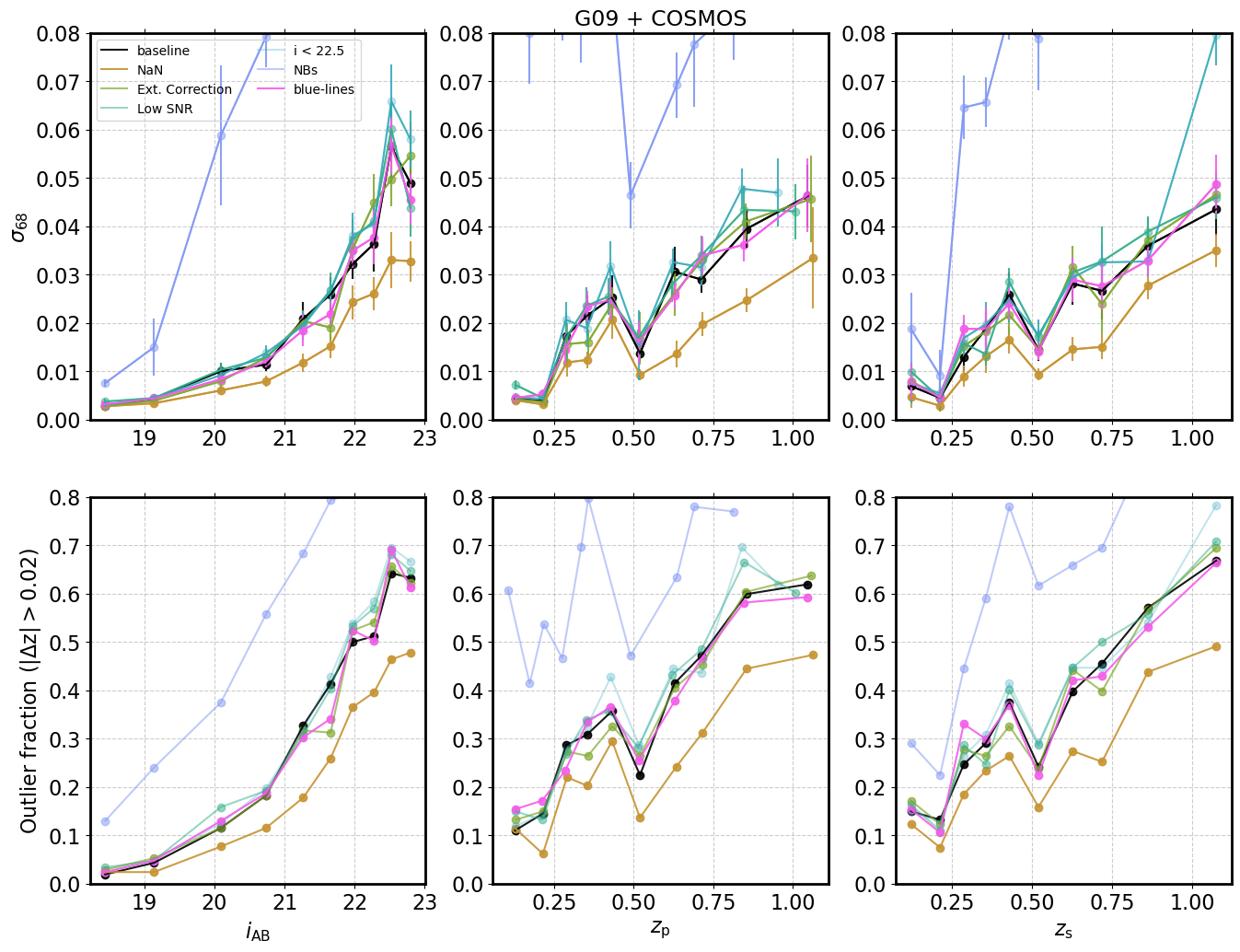}
    \caption{The upper row depicts trends and variations in the measurement of $\sigma_{68}$, while the lower row shows the fraction of outlier values. The distribution is divided into ten bins, each containing an equal number of objects, based on the magnitude in the $i$ band, $z_{\rm p}$, and $z_{\rm s}$, positioned in the left, centre, and right columns, respectively. These trends represent the findings from seven tests, summarised in Table~\ref{tab:studies}, implemented on the W1+W3 and G09+COSMOS baseline samples.}
    \label{fig:validation_set}
\end{figure*}

The comparison between refinement tests is shown in Fig.~\ref{fig:validation_set}, in which we show the trends of $\sigma_{68}$ and the fraction of outliers as a function of apparent magnitudes in the $i$-band, $z_{\rm p}$, and $z_{\rm s}$ for sets of models trained with different galaxy samples. Two baseline samples are taken, one combining galaxies from the W1 and W3 fields of CFHTLenS and the other galaxies from the G09 and COSMOS fields of KiDS plus VIKING. These two samples are adapted to the specific conditions of each test, and each model is trained with them. The trends are obtained with the baseline samples, i.e., without any restriction.

\vspace{0.2 cm}
We now comment on the test in the same orders as they were presented in Sect. 3.3: 

\begin{itemize}

\item {\it CV}:
The CV test revealed that galaxies in the W1, W3 fields, and their combination, do not show a degradation in the precision of $z_{\rm p}$ when including measurements from different filters with similar sensitivities in the $i$-band of the CFHTLenS catalogue. Therefore, the differences between the measurements from two filters are not statistically significant. Consequently, we can obtain more comprehensive catalogues, such that the measurements of both instruments are indistinguishable within the precision limits of the zero point.

\item {\it Infer missing NBs}:
Inferring missing data in the NBs increases the sample size by approximately 35\% in W1 and W3, and around 20\% in G09 and COSMOS. The results suggest that inferring missing data in this way is an effective strategy for improving catalogue completeness without significantly affecting the performance of redshift determination. In fact, this is the test that shows the best performance in both W1+W3 and G09+COSMOS.

\item {\it Low SNR}:
It can be seen that sample loss due to low signal-to-noise ratio in the BBs is not significant in any field, accounting for around 1\% of the sample. Therefore, the performance in these samples is very similar to that obtained with the baseline samples.

\item {\it $i < 22.5$}:
The analysis of galaxies limited to apparent magnitudes brighter than $i_{\rm AB} < 22.5$  indicates good precision, as in the other tests, but shows a decrease in precision for a sample of galaxies that include magnitudes close to 23 and at high redshifts, compared to the other data refinement tests. This slight decrease in precision is a little more noticable in the G09 and fields.

\item {\it Ext.Correction}:
We find that the Galactic latitude of the fields does not have any effect on the precision of $z_{\rm p}$ provided that the extinction correction in Eqn.~\ref{eq:3} is applied. 

\item {\it NBs}:
In the analysis focused on using only the NB in the photometric redshift estimation, in the W1+W3 sample, a reduction in precision is observed with  increasing $i$-band magnitude and redshift, and this more pronounced than in other tests. However, the model performance remains similar to the other cases at low redshifts and for bright galaxies. This suggests that for these two fields, ignoring the BB does not significantly affect the precision in $z_{\rm p}$ for galaxies with $i < 20$, as there is a high SNR in the NBs. For the G09+COSMOS sample, we find poor performance for both faint and bright galaxies, as well as for close and distant galaxies. This finding underlines the importance of the number of examples with which the network is trained, since the training sample in this field is smaller than that for W1+W3, and the NB fluxes have the same SNR.

\item {\it Blue-lines}:
The performance of {\sc DEEPz} does not show any significant changes when the detailed SED features provided by NBs in the blue wavelength range are omitted, so they may not be necessary to estimate $z_{\rm p}$ in the W1, G09 and W3 fields.
\end{itemize}

Since the model trained with the full NB sample has the best performance compared to other tests in W1+W3 and G09+COSMOS, we have decided to adopt this methodology as the main one for determining the photometric redshift. That is, the model will receive as input galaxies with information from 40 NBs + BBs, with the number of the latter depending on the field. In the event that a galaxy has at least ten missing measurements in NBs, a quadratic fit on BBs will be made to infer these measurements. 

To improve the completeness of the photometric redshift catalogues, we include the methodology of models trained with galaxies with imputation in the $i$-band of the CFHTLenS catalogue, i.e., the CV samples and the models trained with galaxies with low SNR. Since these two models show a similar precision to that obtained with the baseline model in the trends of $\sigma_{68}$ and the outlier fraction for weak and distant galaxies in each field or their combination.

\subsection{Effect of spectroscopic surveys}\label{subsec:effect_of_spectroscopic_surveys}

\begin{figure}
    \centering
    \includegraphics[width=1\linewidth]
    {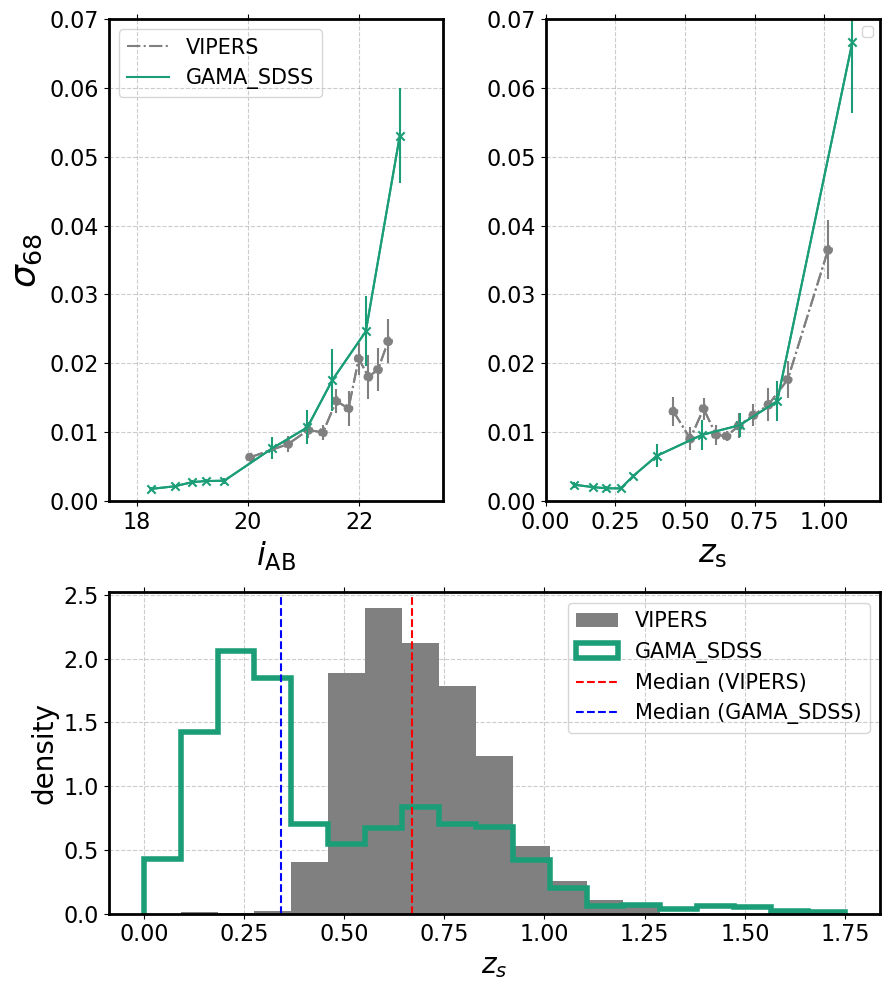}
    \caption{The upper panels show trends in the measurement of $\sigma_{68}$ with respect to $i$ (left panel and $z_s$ (right panel). There are separate curves for the results using the the VIPERS and GAMA\_SDSS redshifts. The lower panel shows the spectroscopic redshift distribution from the VIPERS and GAMA\_SDSS surveys in the W1 field.}
    \label{fig:samples_1_bines_i}
\end{figure}

When spectroscopic catalogues present non-uniform distributions in $z_{\rm s}$, contain sources of variable brightness, data affected by contamination by absorption lines or exposure time limitations,  the recovered $z_{\rm p}$ will reflect these conditions. Therefore, in our analysis, we also consider the origin of the redshifts in the study, since in each field at least the combination of two spectroscopic catalogues is used, and the ranges and statistics of the distributions of $z_{\rm s}$ of these differ, as illustrated in the upper panel of Fig.~\ref{fig:samples_1_bines_i}. We compare the dominant redshift distributions of the baseline sample in the W1 field, VIPERS and GAMA\_SDSS. The upper panels show the trends of $\sigma_{68}$ with respect to $i$ and $z_s$, respectively.

The performance comparison between the VIPERS and GAMA\_SDSS samples was carried out with the same number of sources from the baseline sample in the W1 field, taking the largest possible number of examples from each survey, resulting in samples of 1760 galaxies.

The $\sigma_{68}$ trends in the W1 area, separated by the VIPERS and GAMA\_SDSS samples, show slight differences for faint and distant galaxies. VIPERS has higher precision than GAMA\_SDSS. Although both surveys include measurements for this type of galaxy and were used in the model training, the number of examples from each survey is different in these ranges.

Finding this dependency in the spectroscopic survey data indicates that there will be better precision in the photometric redshifts for galaxies with the characteristics of the survey with the majority of examples.

\subsection{Comparison between \textsc{DEEPz} and \textsc{BCNz2}}\label{subsec:deepzVSbcnz}

We present the $z_{\rm p}$ of the W1 and G09 baseline test samples obtained with models trained with W1+W3 and G09+COSMOS samples, respectively, along with $z_{\rm p}$ obtained in \citet{Navarro_2023}, which use the method {\sc BCNz2}.

\begin{figure*}
    \centering    
    \includegraphics[width=0.49\linewidth]{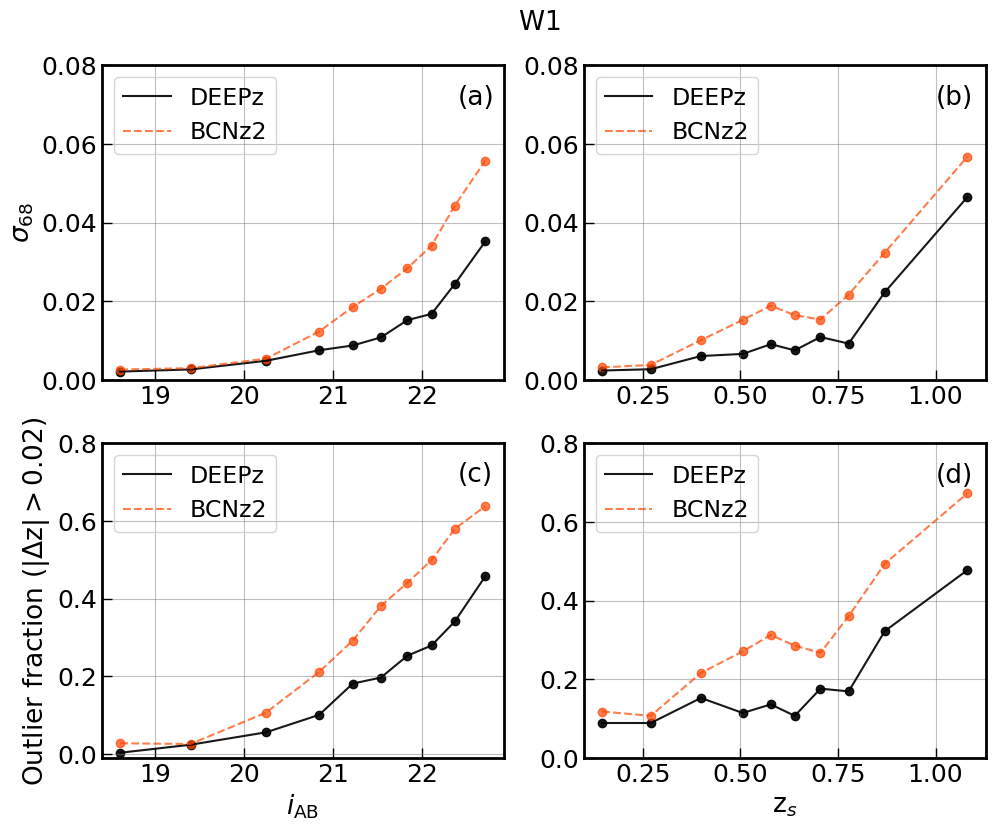}
    \includegraphics[width=0.49\linewidth]{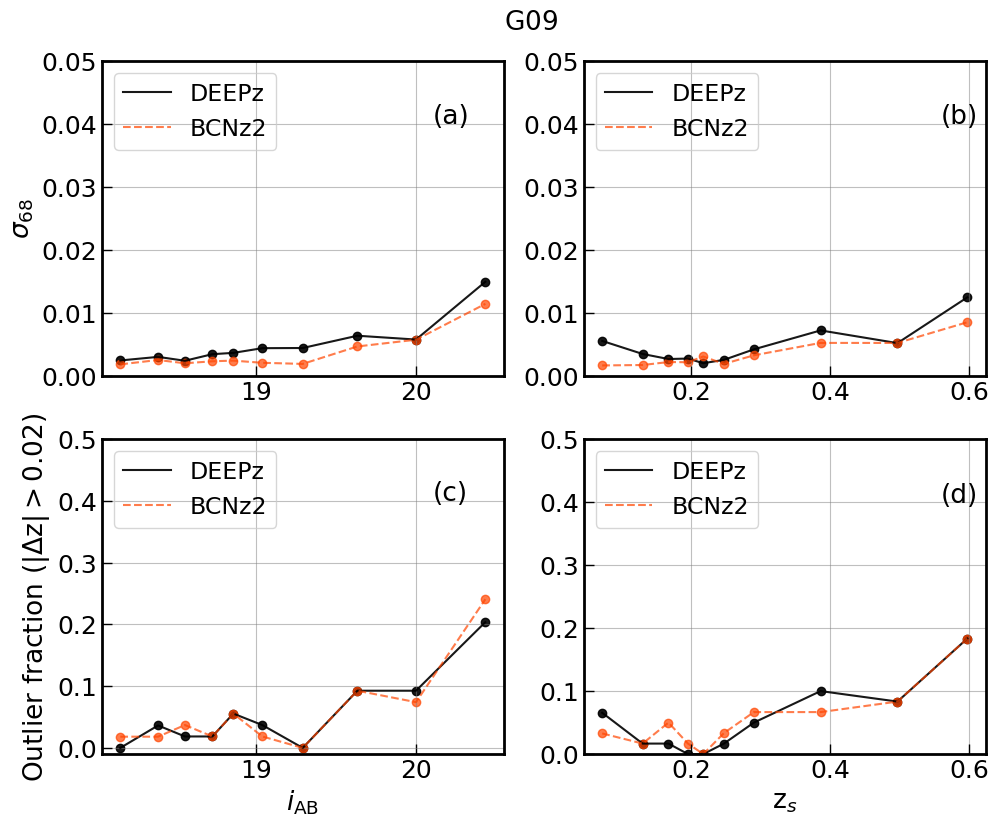}
    \caption{Comparison of the $\sigma_{68}$ values (top row)  and outlier fractions (bottom row) for the W1 and G09 baseline samples obtained using the {\sc DEEPz} and {\sc BCNz2} methods. Columns (1) and (2) correspond to the implementation of the model trained with the W1 sample with missing NB values inferred. Columns (3) and (4) correspond to the implementation of the model trained with the G09+COSMOS sample with missing NBs flues inferred. Upper panels (a) and (b): $\sigma_{68}$ for {\sc DEEPz} and {\sc BCNz2} plotted against $i$-band magnitude for (a) and redshift for (b). Lower panels (c) and (d): Fraction of outliers with $\Delta z > 0.02$ plotted as a function of $i$-band magnitude for (c) and redshift for (d).}
    \label{fig:deepz_BCNz2}
\end{figure*}

In Fig.~\ref{fig:deepz_BCNz2}, we show the trends of $\sigma_{68}$ and the fraction of outliers with respect to the apparent magnitude in the $i$-band and the spectroscopic redshift obtained with {\sc BCNz2} and {\sc DEEPz} for the W1 and G09 baseline samples.

The measurements of $z_{\rm p}$ in the W1 and W3 fields indicate that the $\sigma_{68}$ metric, for both {\sc DEEPz} and {\sc BCNz2}, shows the expected behaviour: a decrease in precision and an increase in the fraction of outliers moving to fainter galaxies and higher spectroscopic redshifts. The performance of the two methods is similar for bright galaxies up to an apparent $i$-band magnitude of approximately 20 and for nearby galaxies up to a redshift of around 0.3. However, beyond these values, {\sc DEEPz} is more accurate than the template method {\sc BCNz2}, reaching a difference in $\sigma_{68}$ of 0.02 (1 + $z_{\rm s}$) between the two methods for faint galaxies with an apparent magnitude in the $i$-band of 23. 

This observed trend between {\sc DEEPz} and {\sc BCNz2} on the W1 baseline sample using the model resulting from the combination of the W1 and W3 fields is consistent with that found in the individual studies of the W1 and W3 fields, for any data refinement test. It is also consistent with the measurements in the COSMOS field, where the methods are applied to samples with a range of apparent magnitude limits in the $i$-band in [20 - 22.5]. However, this behaviour is completely different in the case of bright galaxies in the $i$-band, i.e., for the G09 baseline sample, the {\sc BCNz2} template method is more accurate than {\sc DEEPz} in all cases. However, when a model is implemented with the combination of galaxies from the G09 and COSMOS samples on the G09 baseline sample, {\sc DEEPz} shows an improvement in precision for bright galaxies but not larger than that for {\sc BCNz2}. These results show that the precision of {\sc DEEPz} increases as the number of examples in the training sample increases, as well as when the number of BB filters increases, as demonstrated in the COSMOS field \citep{Eriksen_2020}.

\begin{figure}
\centering
\includegraphics[width=0.9\linewidth]{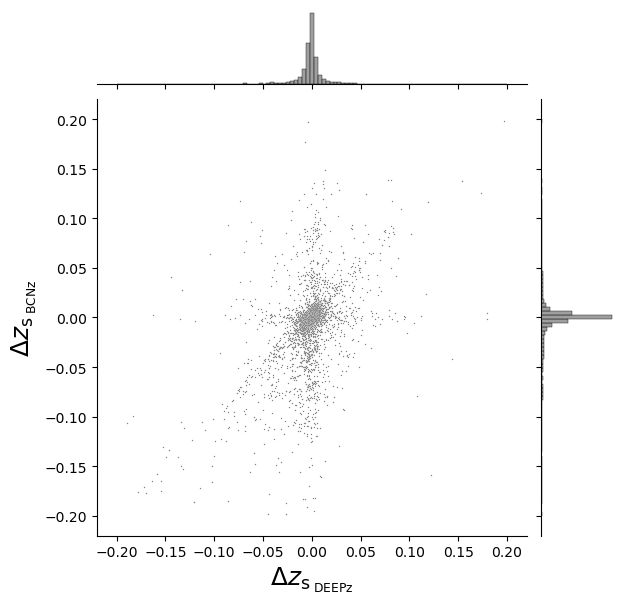}
\includegraphics[width=0.9\linewidth]{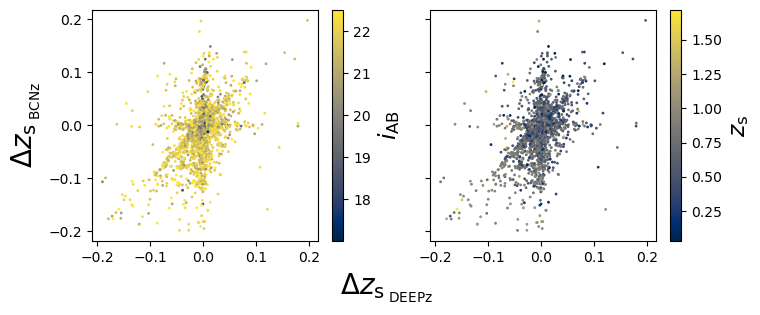}
\caption{Scatter plots of errors on the x-axis for the {\sc DEEPz} method and on the y-axis for the errors obtained when determining $z_{\rm p}$ using the {\sc BCNz2} method. {\it Top panel}: Scatter plot of errors for each method along with a histogram showing the distribution of points along each axis. {\it Bottom left panel}: Scatter plot of errors for each method colour coded according to the apparent magnitude in the $i$ band. {\it Bottom right panel}: Scatter plot of errors for each method colour coded according to $z_{\rm s}$.}
\label{fig:errores_deepz_bcnz}
\end{figure}

Since we have an estimate of the error behaviour of each method according to $i$ and $z_{\rm s}$, we wonder if there is any correlation in the errors of each method, i.e., if the two methods tend to underestimate or overestimate the redshift in the same way. Fig.~\ref{fig:errores_deepz_bcnz} shows the comparison of errors $\Delta z$ from each method, categorised by apparent magnitude in the $i$-band and $z_{\rm s}$ using a colour code. For the baseline sample, we find that the two methods tend towards the perfect case, i.e., if both methods determined $z_{\rm p}$ identical to $z_{\rm s}$, all points would be concentrated at a single point in the centre of the plot. In the scatter plots, we see that the errors of {\sc DEEPz} and {\sc BCNz2} correlate linearly, meaning that there is a set of galaxies for which both methods determine $z_{\rm p}$ with the same level of precision. This is to be expected as both methods use the same photometry. Additionally, on the scatter plots, we observe two patterns, one horizontal and one vertical, located at zero on each axis. These patterns indicate that there are cases where one method is more accurate than the other. However, we find less dispersion overall for the {\sc DEEPz} method. Additionally, we observe that both methods decrease in precision when the apparent magnitude in the $i$-band and the redshift $z_{\rm s}$ increase. 

\subsection{Catalogues}\label{subsec:catalogues}

Based on the results obtained in each field and the study regarding observational effects, as well as the performance comparison between {\sc DEEPz} and the {\sc BCNz2} template method presented in \citet{Navarro_2023}, we have generated catalogues of photometric redshifts in the W1, W3, and G09 fields. In all three fields, we have chosen to utilise the non-NaN model, which uses a training set with inferred values for misssing NBs obtained using a fit to the measurements in the $r$, $g$ and $i$-bands. 

We applied the CV and Low SNR models as additional strategies to enhance the completeness of the catalogues. In the first case, we combined the galaxies belonging to W1 and W3 that have measurements with the second filter in the $i$-band and galaxies with low SNR in both the W1 and W3 fields.
The resulting $z_{\rm p}$ catalogue in the W1 field has 388\,375 galaxies, of which 86 were assigned a $z_{\rm p} = 0$, which means the model could not determine a value. After excluding these, a total of 388\,289 galaxies had a photometric redshift estimated with a mean $z_{\rm p} = 0.59$. In the W3 field, 779\,935 photometric redshifts were obtained with an average $z_{\rm p} = 0.58$, excluding 172 galaxies with $z_{\rm p} = 0$. 
Lastly, in the G09 field, the photometric redshifts of 490\,617 galaxies were estimated, with an average $z_{\rm p} = 0.60$, after discarding 31 galaxies with $z_{\rm p} = 0$.



\section{Close galaxy pairs}\label{close_galaxy_pairs}

To study the performance of $z_{\rm p}$ obtained here in a specific application, the identification of close galaxy pairs are analysed. From the study of the interaction between the members of these systems, it can be inferred how their physical properties are affected \citep{Toomre_1972, Mesa_2014}. The identification of pairs of close galaxies in the W1 and W3 fields and the estimation of their properties are carried out following the methodology described in \citet{Gonzales_2023}. We compare our results to those obtained in the same study, where the photometric redshifts were determined using the {\sc BCNz2} method.

\citet{Gonzales_2023} analyse two samples, one considering the entire galaxy catalogue in each field ({\it total sample}) and the other after applying a quality cut using the photometric redshift values obtained with {\sc BCNz2} ({\it gold sample}). Here, the comparison of the performance of $z_{\rm p}$ in the identification of pairs of galaxies is performed on the {\it total sample}. This sample matches 99.65\% of our catalogue with the {\sc BCNz2} sample in the W1 field and 97.05\% of the {\sc BCNz2} sample in the W3 field.

The identification of pairs of galaxies is carried out using the algorithm described in \citet{Rodriguez_2020a}, which applies criteria regarding the projected distance between galaxies ($r_{\rm p} < 50$ kpc), the difference in projected velocity ($\Delta V < 3500$ \rm{km/s}), and isolation. Despite the difference in the number of galaxies used to generate pairs, the number of pairs of galaxies identified is similar in both fields. In the W1 field, 656 pairs were detected through $z_{\rm p}$ determined by {\sc DEEPz} ($z_{\rm DEEPz}$) and 637 pairs with $z_{\rm p}$ from {\sc BCNz2} ($z_{\rm BCNz2}$). In the W3 field, 1\,521 pairs were found with $z_{\rm DEEPz}$ and 1,627 pairs with $z_{\rm BCNz2}$. Additionally, as shown in Table~\ref{tab:mass_deepz_bcnz}, we also find that the number of galaxy pairs obtained using both catalogues agree in the different samples defined according to the absolute magnitudes in the $r$-band of the pairs, defined as $M_r^{\rm p} = -2.5 \log (L_1 + L_2)$, where $L_2$ and $L_1$ represent the luminosities in the $r$-band of the fainter and brighter galaxies in the pair system, respectively. We also found similarity in the luminosity ratio between the members of the pairs ($L_2/L_1$), the redshift of the pair, and the classification of the pairs according to the colour-magnitude diagram into red and blue categories, as set out in \citet{Gonzales_2023}.

\begin{table}
    \centering
    \begin{tabular}{lrr}
        Sub-sample                    &  {\sc DEEPz} & {\sc BCNz2}\\
        \hline
        \hline
        all pairs                     &  2,157   & 2,282\\
        $M_{\rm r}^{\rm p} <  -22.5$      &  320      & 338\\
        $M_{\rm r}^{\rm p} >= -22.5$      &  1,837   & 1,944\\
        $z_{\rm p} < 0.4$                 &  835      & 852\\
        $z_{\rm p} < 0.4$                 &  1,322   & 1,430\\
        $L_2/L_1 < 0.5$               &  1,221   & 1,251\\
        $L_2/L_1 >= 0.5$              &  936      & 1,031\\
        blue pairs                    &  1,063   & 1,103\\
        red pairs                     &  1,094   & 1,179\\
        \hline
    \end{tabular}
    \caption{Number of close galaxy pairs resulting from the identification with {\sc DEEPz} or {\sc BCNz2}  in total samples and the subsample, selected according to their absolute magnitude in the $r$-band ($M_r$), the pair redshift ($z_{pair}$), the luminosity ratio between the satellite and primary galaxy ($L_2/L_1$), and the colour.}
    \label{tab:mass_deepz_bcnz}
\end{table}

Among the estimated properties, the mass was determined using weak gravitational lensing stacking techniques in the subsets already defined (see Table~\ref{tab:mass_deepz_bcnz}). Through stacking techniques, the signal-to-noise measurement of the lensing effect is improved by artificially increasing the source galaxy density, from which the lens parameters are derived. This allows the calculation of the posterior probability distribution, which in this case refers to the probability distribution of the mass of the galaxy pair. It is obtained by considering Bayesian techniques, where prior information about the parameters (the prior distribution) is combined with the information provided by the data (the likelihood) to obtain the posterior distribution. The posterior density distributions were analysed, considering the physical properties described in Table~\ref{tab:mass_deepz_bcnz}. Additionally, the analysis includes the results obtained in \citet{Gonzales_2023} through the {\sc BCNz2} method on the {\it total sample}. 

\begin{figure}[ht]
    \centering
    \includegraphics[width=\linewidth]{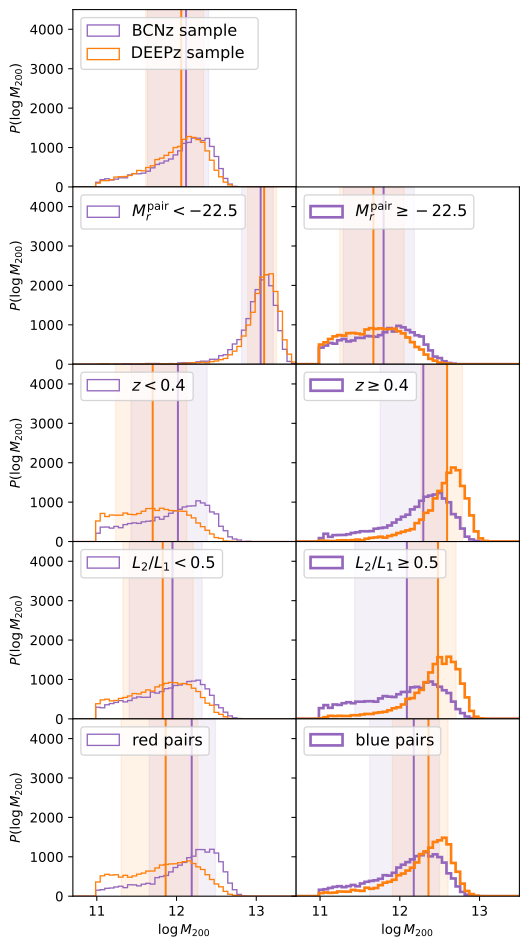}
    \caption{Posterior density distributions of the fitted log $M_{200}$. Each row shows the distributions for the pairs selected from the {\sc BCNz2} and {\sc DEEPz} methods in purple and orange, respectively. The selection cuts according to pair properties are shown using thicker and narrower lines as referred to in the legends. Vertical lines indicate the median values and the shaded regions enclose 68 per cent of the distributions corresponding to the errors. In each panel, the stacked pair sub sample described in Table~\ref{tab:mass_deepz_bcnz} is specified.}
    \label{fig:mass_pair}
\end{figure}

As seen in Fig.~\ref{fig:mass_pair}, the posterior density distributions for the {\sc DEEPz} and {\sc BCNz2} methods are shown for the samples based on magnitude, colour, redshift, and luminosity ratio criteria. For samples selected under magnitude and colour criteria, we found similar trends and values. The probability profiles for pairs with $z < 0.4$ and red pairs show differences, such that the pairs identified with DEEPz that are close and red are less massive compared to {\sc BCNz2}. For pairs with $z \geq 0.4$ and satellite galaxies that meet $L_2/L_1 \geq 0.5$, we found that DEEPz pairs are more massive compared to {\sc BCNz2}. Nevertheless, the error intervals of the medians of all distributions overlap. Consequently, the mass estimates for both samples are statistically equivalent.

While the samples of {\sc DEEPz} and {\sc BCNz2} pairs are compatible in terms of number and mass, even when selecting subsamples, we find that only 249 are common pairs, meaning the central and satellite galaxies of the pair coincide. This amount represents approximately 40\% of the total number of pairs obtained with $z_{\rm DEEPz}$.

Regarding complementary galaxy pairs, which are pairs that do not match in the identifications made through $z_{\rm DEEPz}$ and $z_{\rm BCNz2}$, we find that 250 galaxy members are present in both catalogues. Of these 250 galaxies, it was observed that central galaxies in {\sc DEEPz} are identified as satellites in {\sc BCNz2} in three cases, and vice versa, satellite galaxies in {\sc DEEPz} are classified as central in {\sc BCNz2} in three instances. Despite having the same number, they are not related as pairs; in other words, the central galaxy in one catalogue does not become a satellite in the other. Therefore, the similarity in number and properties results from the identification algorithm, which links close galaxies that are likely in the same environment, implying that they have similar physical properties and also inherit the global properties of the sample used for their identification.

\begin{figure}[ht]
    \centering
    \includegraphics[width=0.9\linewidth]{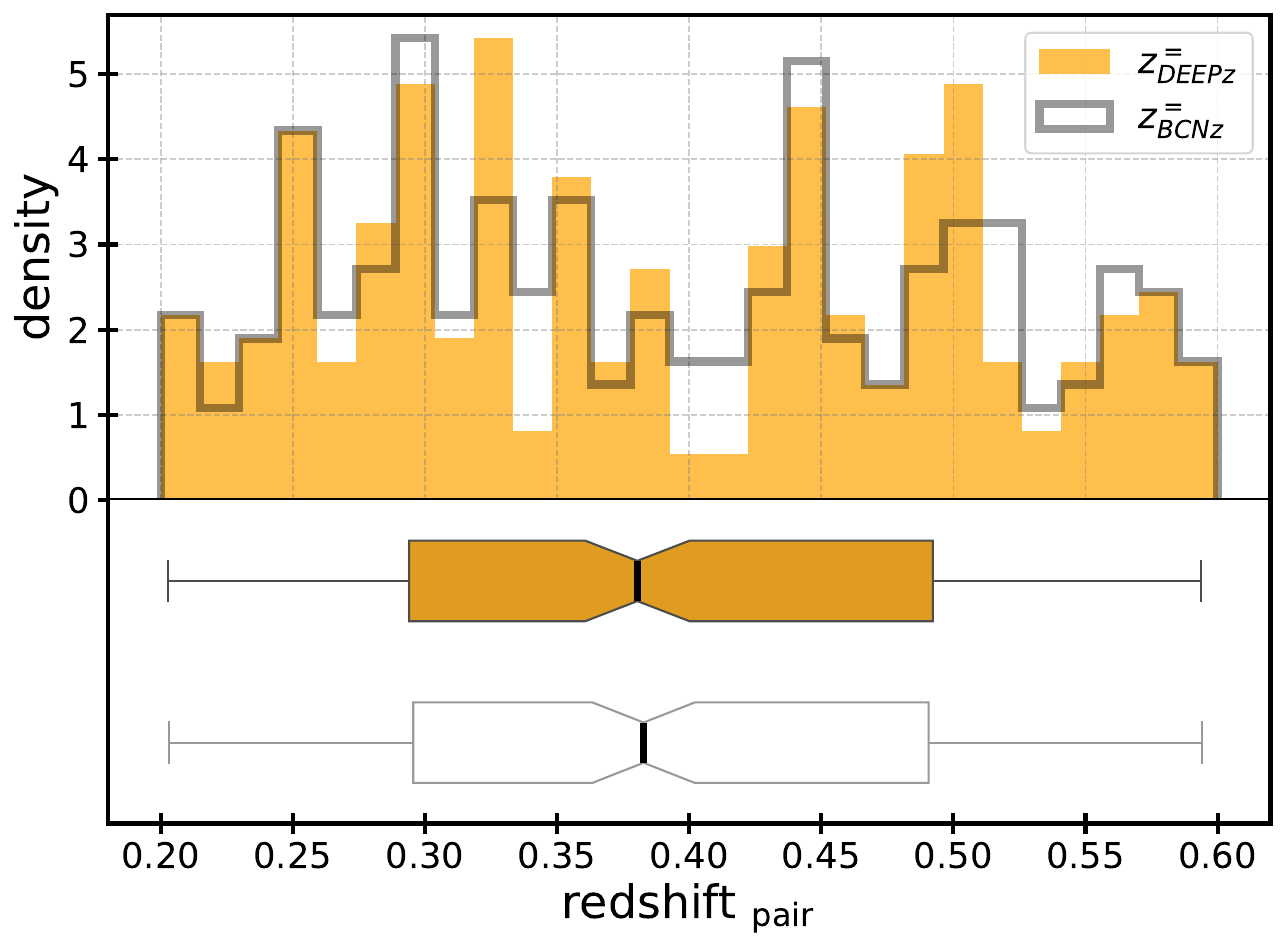}
    \includegraphics[width=0.9\linewidth]{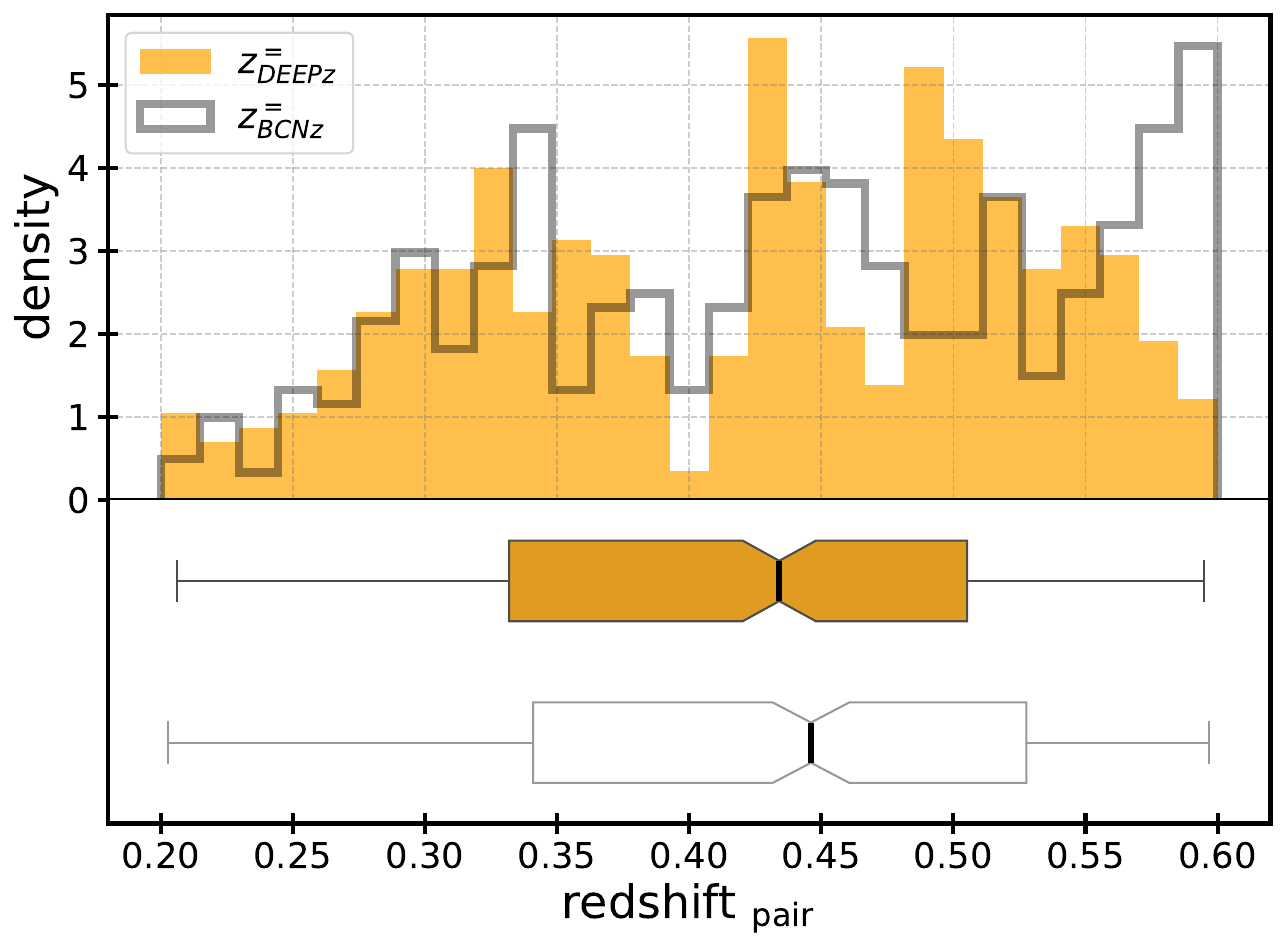}
    \caption{Histograms and box plots are used to compare the distributions of the photometric redshifts of close galaxy pairs. Orange represents the sub samples of pairs determined usinge $z_{\rm DEEPz}$, and black represents sub samples of pairs determined with $z_{\rm BCNz2}$, respectively. {\it Upper panel}: distributions correspond to the samples: ${\rm DEEPz}^{=}$, ${\rm BCNz2}^{=}$. {\it Lower panel}: distributions correspond to the samples: ${\rm DEEPz}^{\neq}$, and ${\rm BCNz2}^{\neq}$.}
    \label{fig:z_par_comparacion}
\end{figure}

When comparing the distributions of photometric redshifts of the pairs, represented in Figure \ref{fig:z_par_comparacion} through histograms and box plots\footnote{In the boxplot diagrams, the top and bottom lines of the boxes represent the 25th and 75th percentiles of the distributions, while the wrists of the boxes represent the medians. 
Notches display the confidence interval (95\% confidence level) symmetrically around the medians. When comparing distributions, if the notches of two boxes do not overlap, there is a statistically significant difference between the medians (\citealt*{boxplot78}; \citealt{boxplot14}).
For skewed distributions or small sized samples it might be the case that the CI is wider than the 25th or 75th percentile, therefore the plot will display some ``inside out'' shape. 
The lines extending from the boxes are called whiskers. The boundary of the whiskers is based on the 1.5 interquartile range (IQR) value. The whiskers extend from the bottom/top of the boxes up to the lowest/largest data point that falls within 1.5 times the IQR. The whisker lengths might not be symmetrical since they must end at an observed data point.}, we observe that the common pairs identified with $z_{\rm DEEPz}$ and $z_{\rm BCNz2}$ ($z_{\rm DEEPz}^{=}$ and $z_{\rm BCNz2}^{=}$) tend to have lower values compared to their complementary sets ($z_{\rm DEEPz}^{\neq}$ and $z_{\rm BCNz2}^{\neq}$). This difference is significant, as the confidence intervals of the box plots do not overlap, indicating that the medians of $z_{\rm p}$ of $z_{\rm DEEPz}^{=}$ and $z_{\rm BCNz2}^{=}$ are statistically different from their complements. However, the statistical difference in the medians of $z_{\rm p}$ between common pairs and their complements is likely because the redshifts of the latter are higher due to the increased volume. This is evident as their $z_{\rm p}$ distribution extends across the entire range.

\begin{figure}
    \centering
    \includegraphics[width=1\linewidth]{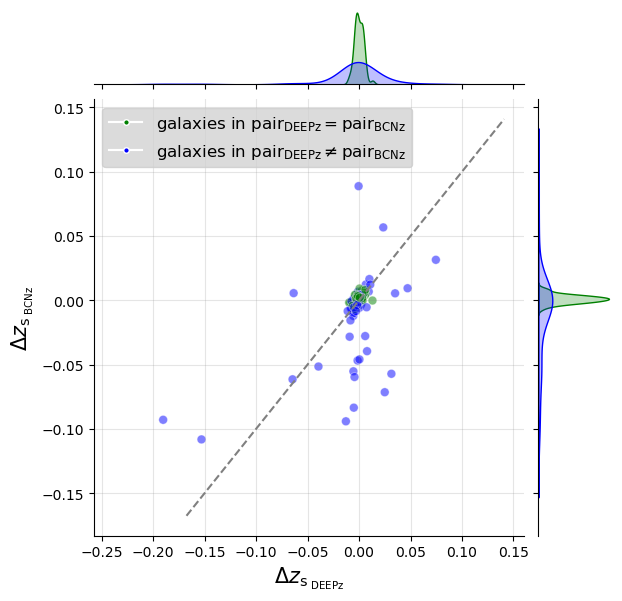}
    \caption{Scatter plot between the errors of {\sc DEEPz} and {\sc BCNz2} along with the density distributions of galaxies in common pairs (${\rm pair}_{\rm DEEPz}$ $=$ ${\rm pair}_{BCNz2}$) and galaxies in non common pairs (${\rm pair}_{\rm DEEPz} \neq {\rm pair}_{BCNz2}$) between the two methods.}\label{fig:errores_par1}
\end{figure}

This behaviour is in agreement with what was observed in the comparison between $\sigma_{68}$ and the fraction of outliers with respect to the apparent magnitude in the $i$-band and $z_{\rm s}$ bands for both methods. The redshifts determined by each method for bright, close galaxies are similar in their trends and precision. Regarding galaxies with apparent magnitude in the $i$-band fainter than 20 and $z_{\rm pair}$ greater than 0.3, the $\sigma_{68}$ measurements between these two methods show differences of up to $\sim$ 0.02 in the estimates of $z_{\rm p}$ for the faintest galaxies with respect to the apparent magnitude in the $i$-band, with {\sc DEEPz} showing higher precision compared to {\sc BCNz2}. Additionally, we find from galaxies with spectroscopic information that the difference between the pairs of each method is due to when one of these methods deviates significantly from the actual redshift value of the member galaxy of the pair, as shown in Fig.~\ref{fig:errores_par1}. Where we observe that the larger the errors in one of the member galaxies of the pair, the more likely the difference between pairs identified with the $z_{\rm p}$ of {\sc DEEPz} and {\sc BCNz2}.

Considering the behaviour between the two methods and the fact that the only difference in identifying close galaxy pairs is the determination of the photometric redshift, it is estimated that {\sc DEEPz} agrees with the purity and completeness ranges of close galaxy pair catalogues, as described in Table 1 of \citep{Rodriguez_2020a}, under the criteria of magnitude $i < 22.5$ and spectroscopic redshift $0.2 < z < 0.6$.

Therefore, we can initially conclude that the photometric redshift catalogues generated with {\sc DEEPz} are effective for the identification of galaxy pairs. Furthermore, through the identification of common pairs and their complements, along with the $\Delta z$ values of the member galaxies obtained with each method, we observe that combining both methodologies can increase the purity of pair catalogues when the same pairs are identified, and also test their performance. However, a more detailed study is needed to correlate the photometric errors of each method with the identification of galaxy pairs and their properties.

\section{Conclusions}\label{sec:conclusion}

PAUS is an optical survey made using the PAUCam instrument on the William Herschel Telescope in La Palma, Canary Islands. The survey employs 40 narrow-band optical filters evenly spaced at 100 $\text{\AA}$ intervals, covering a range from 4500 $\text{\AA}$ to 8500 $\text{\AA}$, and achieving an average spectral resolution of R $\sim$ 65. Focusing on fields such as COSMOS, CFHTLS (W1, W3, W4), and KiDS/GAMA G09, PAUS has covered substantial sky areas: 12.04 ${\rm deg}^{2}$ in W1, 15.7 ${\rm deg}^{2}$ in G09, 22.64 ${\rm deg}^{2}$ in W3, and 1 ${\rm deg}^{2}$ in COSMOS. This extensive coverage supports large-scale cosmological studies and research on galaxy evolution. We present the estimation of the photometric redshift of galaxies in three of the fields observed by PAUS: W1, W3, and G09. Our analysis  includes studies aimed at improving the precision and completeness of the galaxy catalogue by addressing various aspects of the observations and their processing. We compare our \textsc{DEEPZ} photometric redshift measurements with those presented in \citet{Navarro_2023} using the {\sc BCNz2} template method. We implement our photometric redshifts in the identification of close galaxy pairs as a test case. 


For the estimation of $z_{\rm p}$ in each field, a combined approach using simulated and observational data was employed. The observational data included information from individual exposures, combined fluxes from 40 narrow bands (NBs) of PAUS, broadband (BB) data from CFHTLenS in the W1 and W3 fields, and KiDS + VIKING in the G09 field, in addition to spectroscopic redshifts from various studies.


The methodology utilised \textsc{DEEPz} \citep{Eriksen_2020} as the main tool, a code tested in the COSMOS field that includes three neural networks: two autoencoders and a mixture density network. It implements transfer learning, meaning the networks are pretrained with simulated data before being trained with observational data. The training process involves data augmentation by creating co-additions of individual exposures of observations in the 40 NBs. The architecture of the networks is adjusted according to the field or study, analysing the importance of different sets of photometric data to improve the precision of $z_{\rm p}$. The primary metric used to evaluate the precision of photometric redshifts was the centralised scatter, $\sigma_{68}$ (see Eq.~\ref{eq:sigma68}), derived from the distribution of relative differences between spectroscopic and photometric redshifts. Additionally, the fraction of outliers with $|\Delta z| > 0.02$ (see Eq.~\ref{eq:outliers}) was used as a secondary metric.

First, we studied the performance of \textsc{DEEPz} when the models were built using samples from the individual fields: W1, W3, G09, and COSMOS, as well as combinations of W1+W3 and G09+COSMOS. Then, seven tests were conducted regarding data refinement to assess their impact, if any, on the precision of $z_{\rm p}$. Each test generated various datasets that influenced the training of \textsc{DEEPz} and, consequently, the precision of the photometric redshift. More specifically, the tests addressed:

{\it Cross-validation (CV)}:
The impact on redshift precision was evaluated by combining measurements from two different filters ('i.MP9702' and 'i.MP902') of the $i$-band from CFHTLenS. The aim was to validate that there is no significant impact on $z_{\rm p}$ when using either of these two filters.

{\it Impute (non-NaN)}:
To address missing observations in some narrow bands (NBs), missing values were imputed using a quadratic fit based on the $g$, $r$, and $i$ bands. This procedure reduced data loss and allowed the inclusion of galaxies without observations in all NBs, improving the completeness of the galaxy catalogue.

{\it Low Signal-to-Noise Ratio Fluxes (Low SNR)}:
The impact of including sources with at least one broadband (BB) with a low signal-to-noise ratio, classified with a value of 99, was investigated. This allowed generating a catalogue with a larger number of objects.

{\it Bright Galaxies ($i$ < 22.5)}:
The difference in precision was studied by including or excluding faint galaxies. The sample was limited to galaxies with apparent magnitudes less than 22.5 in the target BB of each wide field study.

{\it Galactic Extinction Correction (Ext. Correction)}:
Galactic extinction in the fluxes of NBs and BBs was corrected using Planck Collaboration (2016) \citep{Aghanim_2016} dust maps. This correction was made to account for the effects of differential extinction caused by dust, gas, and stellar density of the Milky Way.

{\it Only NBs (NBs)}:
The utility of using only NBs by eliminating the BBs was investigated. The NBs were normalised by creating an artificial band from the average of the NBs within the wavelength range of the target selection band. This approach allowed studying the importance of more general SED features in the precision of $z_{\rm p}$.

{\it SED in the blue wavelength range (blue-lines)}:
The possibility of saving time by discarding detailed SED information in the blue wavelength range (4550 to 5250 $\text{\AA}$) was explored, based on a univariate analysis showing low correlation in this range. The relevance of the lines identified in the blue region was evaluated concerning the general SED features in all broadband.

Additionally, we studied if there is any dependence on the photometric redshift precision based on the spectroscopic samples used in the training of \textsc{DEEPz}.


We found that the precision of the photometric redshift determined with \textsc{DEEPz} in the wide fields and their combinations is less than $\sigma_{68} = 0.06$ compared to the baseline samples of W1+W3 and G09+COSMOS, which reach apparent magnitudes in the $i$-band of  23 and 22.5, respectively. This behaviour holds, except for the model trained solely with a sample from G09 with $i_{\rm AB}< 20$. However, as expected, in all studies, we observed an increase in the number of outliers as galaxies become fainter and their redshifts grow.

The combination of the W1 with W3 fields and G09 with COSMOS fields produces better results in terms of precision trends (reduced $\sigma_{68}$) and lower outlier fraction with respect to the apparent magnitude in the $i$-band and $z_{\rm s}$, compared to using individual field samples. Regarding the comparison between fields with the same photometric information, there is a systematic predominance of W1 over W3 and G09 over COSMOS. The former is due to differences in sample size, and the latter is attributed to the G09 datasets having spectroscopic information that only considers bright galaxies. Regarding the comparisons between combinations, the W1+W3 combination tends to slightly outperform G09+COSMOS.



Based on the refinement tests and the analysis of the effect of observational samples on the precision of $z_{\rm p}$, we can draw some conclusions:

{\it Instrumental}: Concerning the instrumental refinement tests, the CV study revealed that galaxies in the W1, W3 fields, and their combination, W1 combined with W3, do not show a degradation in the precision of $z_{\rm p}$ when including data from different filters with similar sensitivities in the $i$-band. This also holds true when galaxies are lost due to a low signal-to-noise ratio, as these measurements in each field represent approximately 1\% of the samples.

{\it Properties}: In the analysis focused solely on NB, a reduction in precision was found with increasing magnitudes in the $i$-band and redshifts compared to other studies. However, the model performance remained similar in terms of low redshifts and bright galaxies. This trend suggests that for these specific galaxies, excluding broad bands does not significantly affect the precision of $z_{\rm p}$. This finding highlights the importance of the general SED characteristics provided by broad bands in determining photometric redshifts using the \textsc{DEEPz} method. The results obtained without NBs in the blue wavelength range showed similarities with the low signal-to-noise ratio study, indicating that the detailed characteristics in the SED provided by NBs in the blue wavelength range may not be necessary to estimate $z_{\rm p}$ in the W1, W3, and G09 fields.

{\it Completeness}: The inclusion of inferred NB  data (non-NaN) increased the sample size by approximately 35\% in W1 and W3 and around 20\% in G09 and COSMOS. It also proved optimal for improving the performance of \textsc{DEEPz} compared to the other tests. This suggests that filling in the missing NB values is an effective strategy for generating photometric redshift catalogues with good precision. In G09+COSMOS, this study also shows a superior performance trend compared to other models. However, due to sample sizes, the measurements fall within confidence intervals, implying that the medians in each bin of magnitude and redshift are not statistically different. Despite this, we have consider this approach the best strategy for generating $z_{\rm p}$ catalogues in any of the fields.

{\it Spectral Biases}: We found a dependence of $z_{\rm p}$ precision on the spectroscopic survey used in the training. This gives us an implicit measure that there will be better photometric redshift precision for galaxies with the characteristics of the survey predominant in the number of examples, in our case, VIPERS.

Regarding the comparison between the \textsc{DEEPz} method and the template method \textsc{BCNz2}, we found similar trends in accuracy for both apparent magnitude in the $i$-band and $z_{\rm s}$. However, \textsc{DEEPz} demonstrates a 20 to 50 per cent reduction in $\sigma_{68}$ and outlier fraction for the faintest galaxies ($i=21-23$) compared to \textsc{BCNz2} (see Fig.~\ref{fig:deepz_BCNz2}). We also observed a linear correlation between their errors, with both methods slightly tending to underestimate errors within certain redshift ranges.

Finally, as an example test case, following the methodology of \citet{Gonzales_2023}, where galaxy pairs are identified using $z_{\rm p}$ obtained with {\sc BCNz2}, we have implemented the $z_{\rm p}$ galaxy catalogues determined with {\sc DEEPz} in the W1+W3 fields with a precision of $\sigma_{68} = $ 0.01. Based on the results of \citet{Gonzales_2023} and our own results, we have found a good agreement in the number of identified pairs and their physical properties. However, despite this agreement in number and physical properties, {\sc DEEPz} and {\sc BCNz2} only match 40\% of the total sample of galaxy pairs concerning the {\sc DEEPz} sample, characterised by including photometric redshifts of nearby galaxies. Regarding the 60\% of the pairs identified by {\sc DEEPz} that are not found in the identification by {\sc BCNz2}, we found that this difference is due to the low precision for distant galaxies by both methods, with {\sc BCNz2} having a greater error in these cases. Nevertheless, since all the galaxy pairs identified with redshifts provided by {\sc DEEPz} and {\sc BCNz2} meet the purity and completeness criteria detailed in \citet{Rodriguez_2020a}, we consider that the catalogues generated here are capable of identifying close galaxy pairs and that the combination of both methods improves their purity.

As future work, we note that the precision of {\sc DEEPz} increases with more training examples. {\sc DEEPz} shows biases from the spectroscopic surveys used in its training, and its precision is comparable to or better than {\sc BCNz2} in certain fields. These aspects and biases can be improved by including more spectroscopic information, which is time-consuming. Implementing data augmentation and sampling techniques could potentially solve this issue. This will be crucial to continue with transfer learning to improve {\sc DEEPz} and make the models more general. This, in turn, allows for the expansion of the field of implementation of the catalogues for a more precise understanding of the Universe.

\section*{Acknowledgements}
We would like to thank the Hanyue Guo for the useful comments and suggestions which has helped to improve this paper.

CosmoHub has been developed by the Port d’Informació Científica (PIC), maintained through a collaboration of the Institut de Física d’Altes Energies (IFAE) and the Centro de Investigaciones Energéticas, Medioambientales y Tecnológicas (CIEMAT) and the Institute of Space Sciences (CSIC \& IEEC).
CosmoHub was partially funded by the “Plan Estatal de Investigación Científica y Técnica y de Innovación” program of the Spanish government, has been supported by the call for grants for Scientific and Technical Equipment 2021 of the State Program for Knowledge Generation and Scientific and Technological Strengthening of the R+D+i System, financed by MCIN/AEI/ 10.13039/501100011033 and the EU NextGeneration/PRTR (Hadoop Cluster for the comprehensive management of massive scientific data, reference EQC2021-007479-P) and by MICIIN with funding from European Union NextGenerationEU(PRTR-C17.I1) and by Generalitat de Catalunya.

The PAU data centre is hosted by the Port d’Informació Científica (PIC), maintained through a collaboration of CIEMAT and IFAE, with additional support from Universitat Autònoma de Barcelona and ERDF. We acknowledge the PIC services department team for their support and fruitful discussions. This project has received funding from the European Union’s Horizon 2020 Research and Innovation Programme under the Marie Sklodowska-Curie grant agreement No 734374 and HORIZON-MSCA-2021-SE-01 Research and Innovation programme under the Marie Sklodowska-Curie grant agreement number 101086388. This work was also partially supported by the Consejo Nacional de Investigaciones Científicas y Técnicas (CONICET, Argentina), Agencia Nacional de Promoci\'on Cient\'ifica y Tecnol\'ogica and the Secretaría de Ciencia y Tecnología de la Universidad Nacional de Córdoba (SeCyT-UNC, Argentina). This work has made use of CosmoHub. CosmoHub has been developed by the Port d’Informació Científica (PIC), maintained through a collaboration of the Institut de Física d’Altes Energies (IFAE) and the Centro de Investigaciones Energéticas, Medioambientales y Tecnológicas (CIEMAT), and was partially funded by the “Plan Estatal de Investigación Científica y Técnica y de Innovación” program of the Spanish government. 

M. Eriksen acknowledges funding by MCIN with funding from European Union NextGenerationEU (PRTR-C17.I1) and by Generalitat de Catalunya.
This work has been also partially supported by the Polish National Agency for Academic Exchange (Bekker grant BPN/BEK/2021/1/00298/DEC/1), the European Union’s Horizon 2020 Research and Innovation programme under the Maria Sklodowska-Curie grant agreement (No. 754510).
F. Rodriguez would like to acknowledge support from the ICTP through the Junior Associates Programme 2023-2028.
C.M Baugh acknowledges support from the Science Technology Facilities Council through grant number ST/X001075/1.
A. Wittje is supported by the DFG (SFB 1491). CIEMAT participation is supported by the grant PID2021-123012NB-C42P funded by MCIN/AEI /10.13039/501100011033. 
C. Padilla acknowledges support from the Spanish Plan Nacional project PID2019-111317GB-C32 and PID2022-141079NB-C32. 
P. Renard acknowledges the support by the Tsinghua Shui Mu Scholarship, the funding of the National Key R\&D Program of China (grant no. 2023YFA1605600), the science research grants from the China Manned Space Project with No. CMS-CSST2021-A05, the Tsinghua University Initiative Scientific Research Program (No. 20223080023) and the National Science Foundation of China (grant no. 12350410365).
J. Carretero acknowledges support from the grant PID2021-123012NA-C44 funded by MCIN/AEI/ 10.13039/501100011033 and by “ERDF A way of making Europe”. H. Hildebrandt is supported by a DFG Heisenberg grant (Hi 1495/5-1), the DFG Collaborative Research Center SFB1491, as well as an ERC Consolidator Grant (No. 770935).
M. Siudek acknowledges support by the Polish National Agency for Academic Exchange (Bekker grant BPN/BEK/2021/1/00298/DEC/1), the State Research Agency of the Spanish Ministry of Science and Innovation under the grants 'Galaxy Evolution with Artificial Intelligence' (PGC2018-100852-A-I00) and 'BASALT' (PID2021-126838NB-I00). This work was partially supported by the European Union's Horizon 2020 Research and Innovation program under the Maria Sklodowska-Curie grant agreement (No. 754510).

\section*{Data Availability}
The {\sc DEEPz} photometric redshift catalogues in the W1, W3, and G09 fields will become publicly available on the CosmoHub platform \citep{TALLADA2020100391, 2017ehep.confE.488C} in an upcoming data release at \url{https://cosmohub.pic.es/catalogs/319}. The data is currently available on a reasonable request to the author.

\bibliographystyle{aa}
\bibliography{references}

\appendix
\section{Feature selection}\label{app:appendix_a}

\begin{figure*}
    \centering
    \includegraphics[width=1\linewidth]{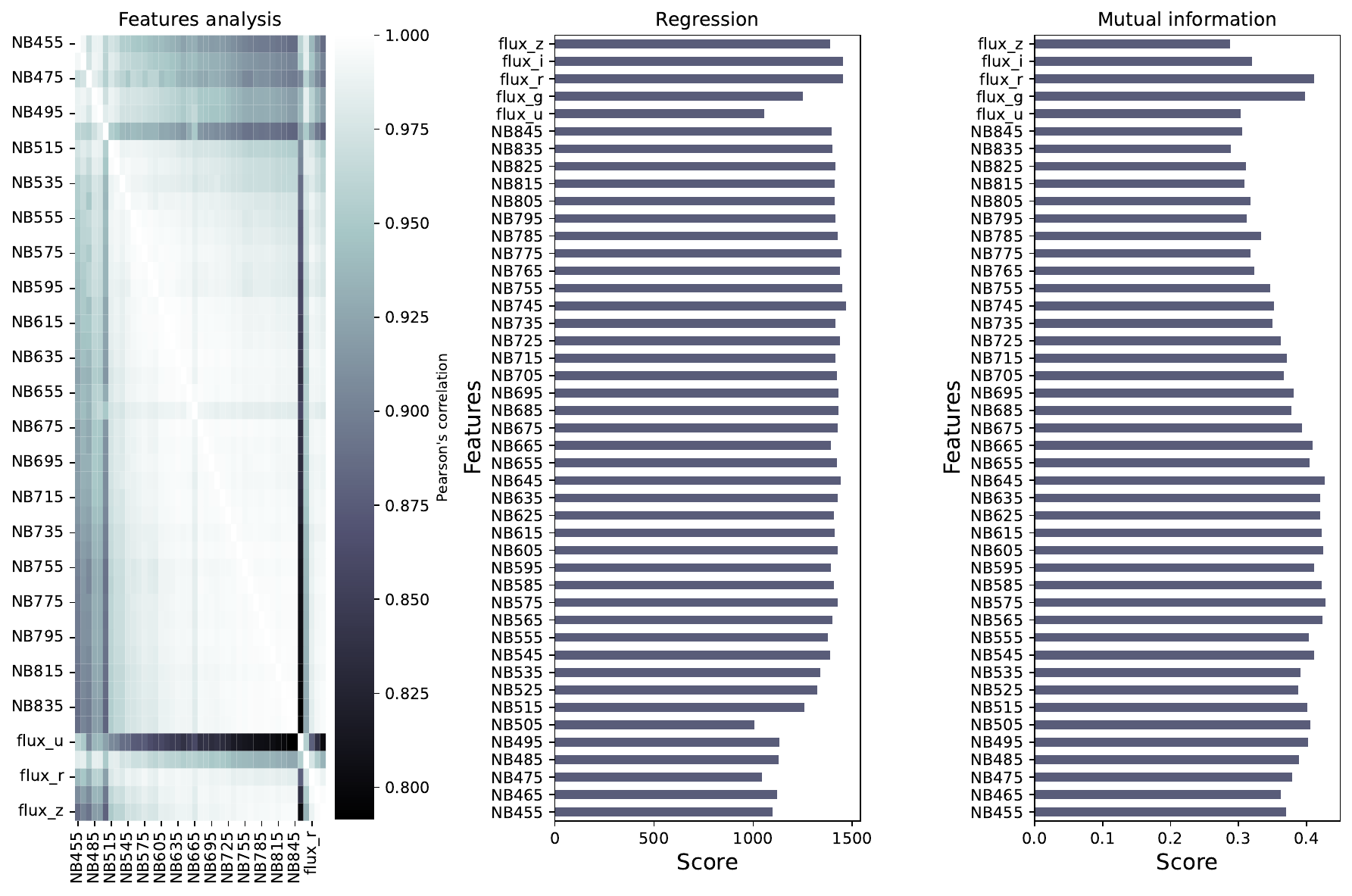}
    \caption{Visualisation of univariate study of bands used in W1 and W3 fields.
    {\it Left panel}: correlation among the bands through Pearson correlation. 
    {\it Central panel and right panel}: Bar chart of the scores of the correlated bands with the spectroscopic redshift bins calculated through MI and regression methods.}
    \label{fig:combined_figures.pdf}
\end{figure*}

One of the most important steps in classifier determination is generating and selecting the features with the highest entropy. To do this, we implemented different statistical methods, particularly unsupervised learning and univariate analysis. In Fig.~\ref{fig:combined_figures.pdf}, we show in the left panel the heat map of the different Pearson correlations between the bands, and in the central and right panels, the scores obtained applying the mutual information and regression methods, respectively. This type of analysis allows us to estimate the importance of the features, i.e., the bands for the analysis or modeling, and to discard irrelevant or redundant ones. Reducing the number of features while preserving the most important information improves the efficiency and interpretation of the models. We observe that the trend of the BBs is similar to the trend of the NBs in terms of importance when studying their relationship with the spectroscopic redshift intervals, so it is interesting to study the necessity of the BBs. Additionally, we note that the regression method assigns a lower score to the bands of blue wavelengths compared to the rest of the bands, so it is interesting to study their importance in determining the $z_{\rm p}$.

\section{Performance for field and refinement data test}\label{app:appendix_b}

\begin{figure*}
    \centering
    \includegraphics[width=0.8\linewidth]{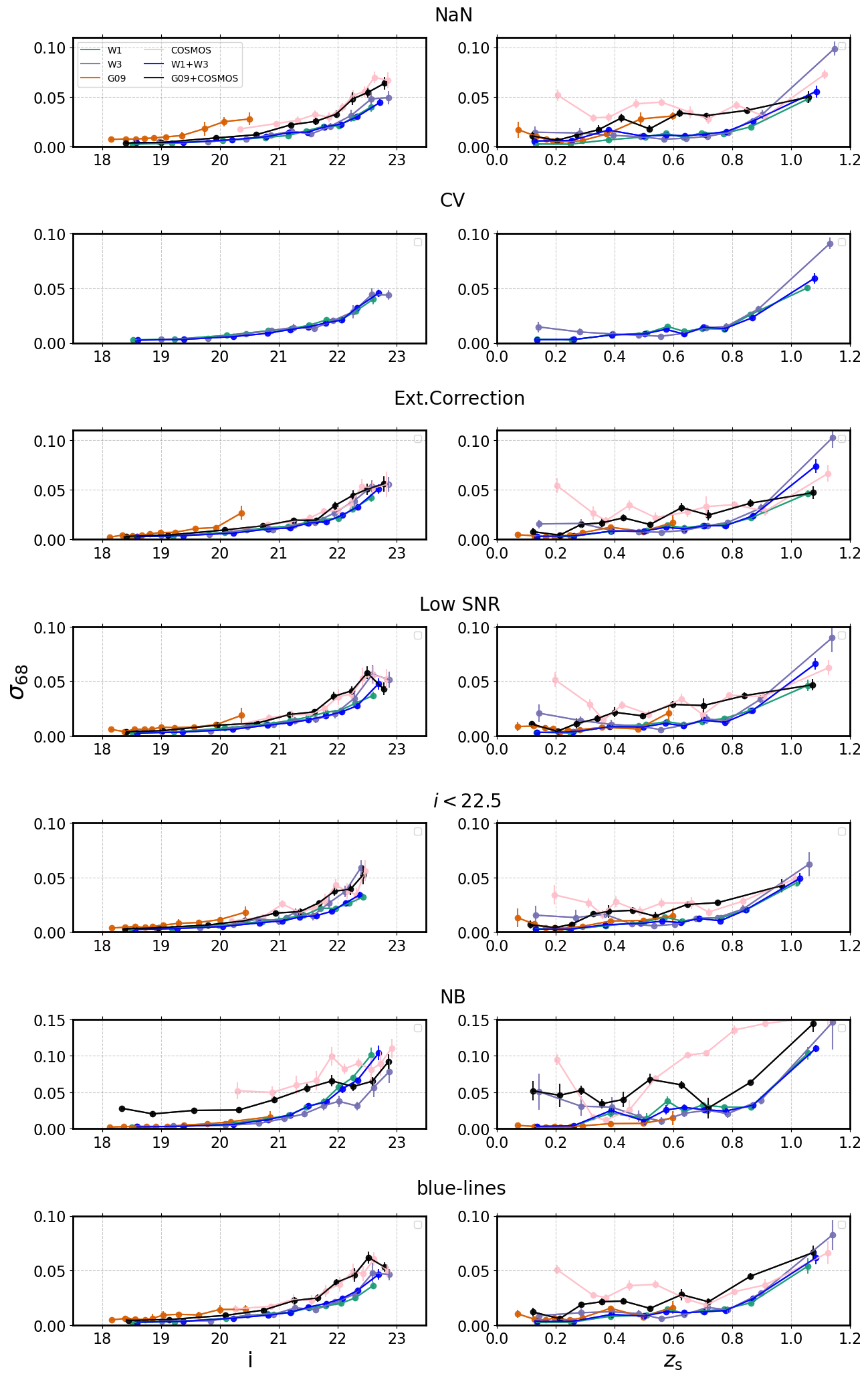}
    \caption{Trend of $\sigma_{68}$ as a function of the $i$-band magnitude and spectroscopic redshift for each field and the combination W1+W3 and G09+COSMOS. Each row corresponds to a data refinement study, hence the validation sets are different.}
    \label{fig:refinement_data_w1_w3_g09}
\end{figure*}

\begin{figure}
    \centering
    \includegraphics[width=1\linewidth]{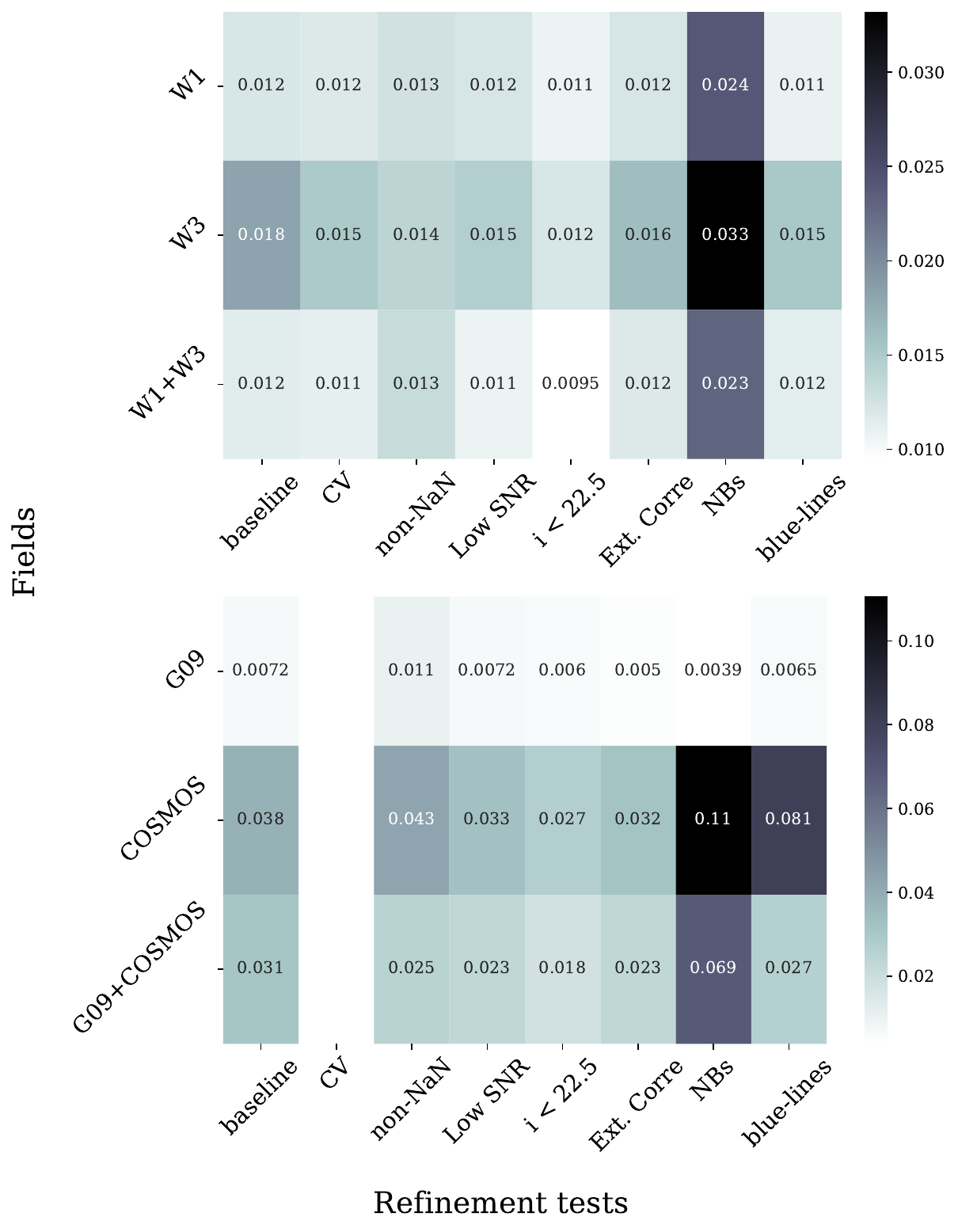}
    \caption{Heatmaps of $\sigma_{68}$ from the validation sample of each study. {\it Top panel}: Galaxy samples in W1, W3, and W1+W3 combined. {\it Bottom panel}: Galaxy samples in G09, COSMOS, and G09+COSMOS combined.
}
    \label{fig:validation_set_map}
\end{figure}

Each row corresponds to a different refinement test for the five samples. In the left column, the results are presented according to 10 $i$-band bins, while in the right column, the results are shown according to 10 $z_{\rm s}$ bins. Each result is linked to the initial training and validation samples within each field and its combination; these samples may increase, decrease, or remain constant in the number of examples, depending on the refinement data test being conducted. Therefore, the intervals in each curve of each figure vary. The different data refinement tests are showed in Fig.~\ref{fig:refinement_data_w1_w3_g09} and are summarised in the heatmap of Fig.~\ref{fig:validation_set_map}, this plots are showing the value of the $\sigma_{68}$ metric for the validation sample.

In general, we find that $\sigma_{68}$ as a function of $i$ and $z_{\rm s}$ varies in the same way as seen in the Sect.~\ref{sec:catalogue} for the baseline samples. However, this behaviour is not observed for the G09 field. This is because the samples from the G09 field have spectroscopic information restricted up to $i = 20.5$ and $z_{\rm s} = 0.6$, i.e., it considers only bright galaxies. Therefore, in this field, lower values are generally observed in the metrics compared to other samples.

\section{Extrapolation to other regions of the sky}\label{app:appendix_c}

\begin{figure*}
    \centering
    \includegraphics[width=1\linewidth]{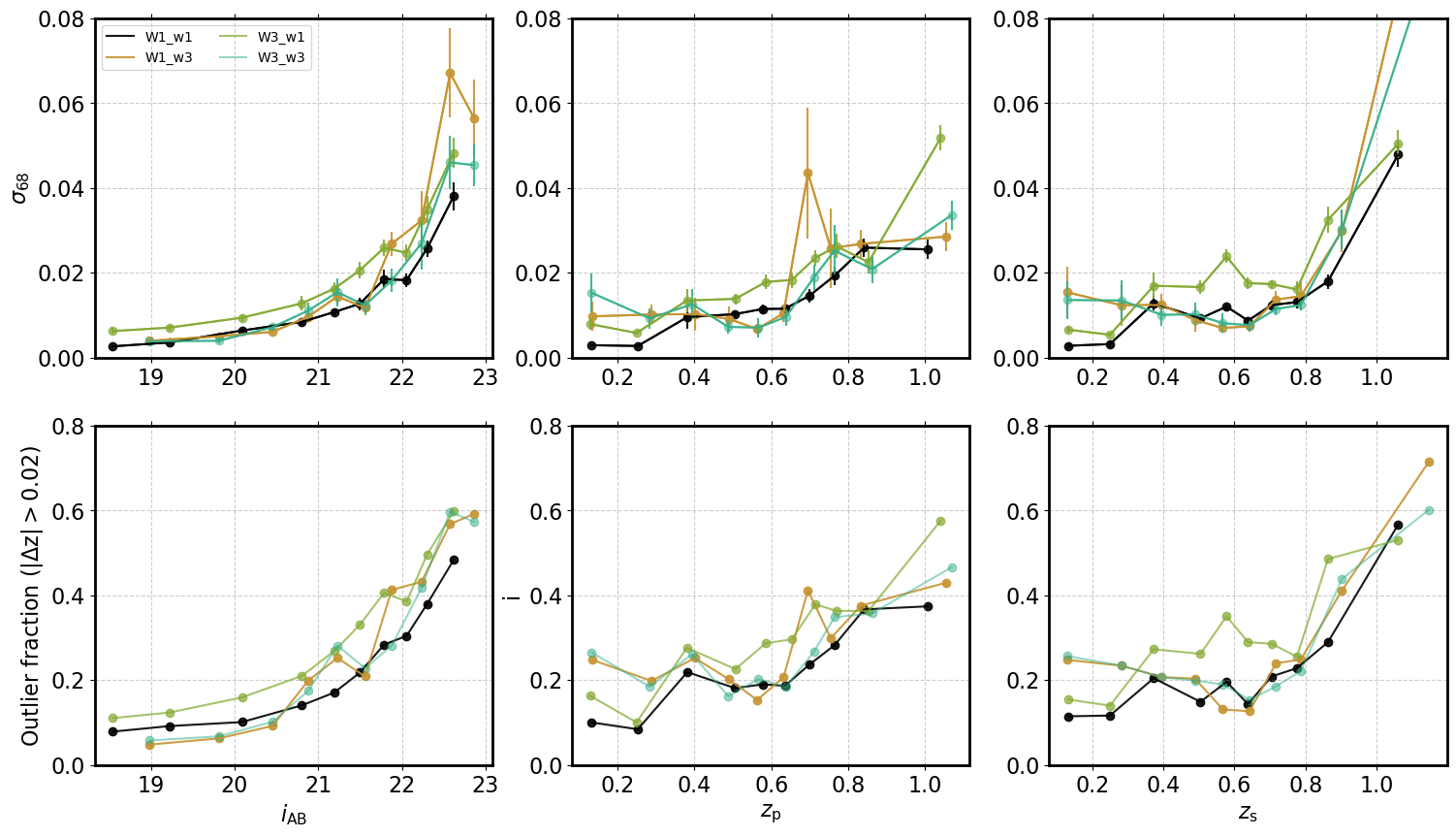}
    \caption{Trends in the measurement of $\sigma_{68}$. The distribution is divided into ten bins, each containing the same number of objects, based on the apparent magnitude in the $i$-band and $z_{\rm s}$, in the left and right columns, respectively. These trends show the precision, in terms of the centralised scatter $\sigma_{68}$, in the W1 and W3 validation sets. The solid line represents when the training and validation samples coincide, and the dashed line indicates training with W1 and validation with W3, or vice versa.}
    \label{fig:field_mix}
\end{figure*}

In this appendix we study the impact of applying {\sc DEEPz} on sky regions not included in the training set. Specifically, we test how the performance changes if being is trained in W1 and validated in W3 or vice versa. In Fig.\ref{fig:field_mix}, you can see $\sigma_{68}$ as a function of the $i$-band magnitude, photo-$z$, and spectroscopic redshift.

The trends shown in Fig.~\ref{fig:field_mix}, when the training and validation samples match, are consistent with those seen in Fig.~\ref{fig:refinement_data_a} and Fig.~\ref{fig:validation_set_map}. A higher precision is observed in the W1 field compared to the W3 field, with a slight deterioration when the validation samples are swapped between these two fields. However, this decrease is more pronounced when DEEPz is trained exclusively with galaxies from the W3 field and applied to galaxies from the W1 field. Based on these trends, it is likely that performance in new fields will vary slightly in new regions.
\end{document}